\documentclass[letterpaper]{article} 
\usepackage{aaai23}  
\usepackage{times}  
\usepackage{helvet}  
\usepackage{courier}  
\usepackage[hyphens]{url}  
\usepackage{graphicx} 
\usepackage{natbib}  
\usepackage{caption} 
\usepackage{appendix}
\usepackage{graphicx}
\usepackage{lineno}
\usepackage{array}
\usepackage{longtable}
\usepackage{multirow}
\usepackage{listings}
\usepackage{color}
\usepackage{graphics}
\usepackage{tabularx}
\usepackage{longtable}
\usepackage{array} 
\usepackage{booktabs}
\usepackage{epstopdf} 
\usepackage{multirow}
\usepackage{pdflscape}
\usepackage{afterpage}
\usepackage{capt-of}
\usepackage{rotating}
\usepackage{algorithmic}
\usepackage{algorithm}
\usepackage[utf8]{inputenc}
\usepackage{placeins}
\usepackage{ragged2e}
\usepackage[export]{adjustbox}
\usepackage{adjustbox}
\usepackage[misc]{ifsym}
\usepackage{notoccite}
\usepackage{xspace}
\usepackage{caption}
\usepackage{subcaption}
\usepackage{amsmath}
\usepackage{amsthm}
\usepackage{cleveref}

\usepackage{newfloat}
\usepackage{listings}
\DeclareCaptionStyle{ruled}{labelfont=normalfont,labelsep=colon,strut=off} 
\floatstyle{ruled}
\newfloat{listing}{tb}{lst}{}
\floatname{listing}{Listing}
\newcommand{\one}{({\em i}\/) \xspace}
\newcommand{\two}{({\em ii}\/) \xspace}
\newcommand{\three}{({\em iii}\/) \xspace}

\crefformat{section}{\S#2#1#3} 
\crefformat{subsection}{\S#2#1#3}
\crefformat{subsubsection}{\S#2#1#3}

\def\eg{\emph{e.g.} \xspace}
\def\etc{\emph{etc.} \xspace}
\def\ie{\emph{i.e.} \xspace}

\newcommand{\pb}[1]{\vspace{0.75ex}\noindent{\bf \em #1}\hspace*{.3em}}

\title{Lady and the Tramp Nextdoor:\\ Online Manifestations of Economic Inequalities in the Nextdoor Social Network}
\author{
    Waleed Iqbal\textsuperscript{\rm 1}, Vahid Ghafouri\textsuperscript{\rm 2, \rm 3}, Gareth Tyson\textsuperscript{\rm 1, \rm 4},\\
    Guillermo Suarez-Tangil\textsuperscript{\rm 2}, Ignacio Castro\textsuperscript{\rm 1}
}
\affiliations{
    \textsuperscript{\rm 1}Queen Mary University of London, UK,\\ \textsuperscript{\rm 2}IMDEA Networks Institute, Spain,\\ \textsuperscript{\rm 3}Universidad Carlos III de Madrid, Spain,\\ \textsuperscript{\rm 4}Hong Kong University of Science and Technology (GZ), China\\
\{w.iqbal, i.castro\}@qmul.ac.uk,
\{vahid.ghafouri, guillermo.suarez-tangil\}@imdea.org,
gtyson@ust.hk
}

\usepackage{bibentry}

\begin{document}

\maketitle

\begin{abstract}
From health to education, income impacts a huge range of life choices.
Earlier research has leveraged data from online social networks to study precisely this impact.
In this paper, we ask the opposite question: do different levels of income result in different online behaviors?
We demonstrate it does. We present the first large-scale study of Nextdoor, a popular location-based social network. We collect 2.6 Million posts from 64,283 neighborhoods in the United States and 3,325 neighborhoods in the United Kingdom, to examine whether online discourse reflects the income and income inequality of a neighborhood.
We show that posts from neighborhoods with different incomes indeed differ, \eg richer neighborhoods have a more positive sentiment and discuss crimes more, even though their actual crime rates are much lower.
We then show that user-generated content can predict both income and inequality.
We train multiple machine learning models and predict both income (R\textsuperscript{2}=0.841) and inequality (R\textsuperscript{2}=0.77).
\end{abstract}

\section*{Introduction}
Income is a critical factor in many aspects of life and a large body of research has examined the effect of income on how individuals interact with the real world~\cite{wu2012neighborhood, dekker2007social, hays2007neighborhood}. 
In this paper, we address the opposite question: do income differences result in differences in online  discourse?
We hypothesize that this is the case~\cite{bernstein1960language}
and that these differences can be leveraged to infer the economic context of a person. 

To test our hypothesis we need user data on online discourse and income.
Online interaction data is abundant, but inferring the income of online users is challenging. 
Mobility data and location-based applications such as Foursquare can be used to infer the residential location of a user and then using official statistics associate, the user with the level of income of the user's neighborhood~\cite{aggarwal2013detection,chorley2016pub}. 
This presents two challenges:
\one the residential location of the user needs to be inferred using heuristics, and \two the data might inform of the user's location but will have limited vantage on the online discourse of the user.

Alternatively, online discourse data (\eg from Twitter or Mastodon) can be used to study the content posted by online users and develop heuristics to infer their income, \eg  using the expected level of income for a given professional occupation~\cite{preoctiuc2015analysis, aletras2018predicting}. 
However, this implies that
\one income inferences rely on heuristics, and
\two there might be selection bias as not all users might post content that can be used to infer their income.

In this work, we overcome those limitations by collecting and analyzing a previously unstudied location-based social network with over 10 million users, \emph{Nextdoor}~\cite{nextdoor-stats}. 
Nextdoor users interact within closed social networks of neighbors, \ie users registered in the same neighborhood. To ensure that users register in the location where they reside, new users have to validate their home addresses using regular (snail) mail. 
This allows us to associate neighbors with the median income of their neighborhood and study whether differences in income are reflected in their online discourse.

We present the first large-scale dataset and analysis of Nextdoor.
We collect 2.6 Million posts from 64,283 neighborhoods in the United States (USA) and 3,325 neighborhoods in the ten most populous cities in the United Kingdom (UK) between November 2020 and September 2021. 
We augment this with  official statistics on population, income, and crime for both countries and at the same level of geographical granularity as the Nextdoor neighborhoods. 

We use our augmented dataset to 
study whether differences in income indeed result in online  discourse differences with the following Research Questions (RQs):

\begin{itemize}
    \item \textbf{RQ1:} Do  neighborhoods with  different income levels differ in how they discuss crime  and the sentiment of their posts?
    \item \textbf{RQ2:} Do  neighborhoods with  different levels of income inequality differ in how they discuss crime and on the sentiment of their posts?
    \item \textbf{RQ3:} Can Nextdoor discussions be used to predict  the income and income inequality of a user?
\end{itemize}

While answering these questions, we conduct a neighborhood-level analysis of our Nextdoor dataset.
We examine how  neighborhoods with different \emph{income levels} 
exhibit different traits in the online user-generated content.
We identify strikingly clear differences.
We find that richer neighborhoods seem more concerned about crime:
The 20\% richest neighborhoods discuss crime 1.6x and 1.47x more than the 80\% poorest counterparts in USA and UK, respectively. 
This is the case even though the incidence of crime is 1.24x and 1.36x higher in poorer neighborhoods in USA and UK.
%
We also find that richer neighborhoods seems to post more positive content:
the sentiment of the text in the posts of the 20\%  richest neighborhoods is 1.37x and 1.42x more positive than in the 80\% poorest neighborhoods in USA and UK.

We then examine \emph{income inequality}.
We determine whether a neighborhood is more or less unequal by comparing its median income with the income of the neighborhoods in its vicinity. 
We then look for differences in the content posted between neighborhoods with different levels of inequality in their vicinity.
Again, we find clear differences between neighborhoods depending on the level of income inequality.
The richest neighborhoods with the most \emph{equal} vicinity, discuss crime more than any other neighborhood, in proportion to the official crime reported.
We also find that the richest neighborhoods with the most \emph{equal} vicinity have a more positive sentiment than any other neighborhood.
The opposite is true for the poorest neighborhoods, where the neighborhoods with the most equal vicinities have the lowest sentiment of all.

We then wonder whether we can \emph{predict} a neighborhood's income based on the text posted.
We show that this is indeed the case. 
We predict the income of a neighborhood exclusively using text features obtained from the posts (up to 0.69 R$^2$ with a Lasso regressor). 
We then experiment with several machine learning models to predict income and inequality using features extracted from Nextdoor. 
We find that we can predict a neighborhood's income and its surroundings' inequality with even a higher degree of accuracy (income best R$^2$: 0.841, inequality best R$^2$: 0.77).  
We find this concerning, as it could facilitate income-based discrimination and  algorithmic surveillance~\cite{zuboff2015big}.

This paper makes the following contributions:
\begin{itemize}
    \item We conduct the first large-scale quantitative analysis of  Nextdoor with 
    2.6 Million posts from 64,283 and 3,325  USA and UK neighborhoods, respectively.
    \item We show that the online content generated by the users reveals socioeconomic factors where a greater income of a neighborhood and its surroundings is associated with more crime-sensitive posting activity and more positive sentiment in the posts.
    \item We demonstrate that the features extracted from the user-generated content can predict both the income (R\textsuperscript{2}=0.841) and inequality (R\textsuperscript{2}=0.77) of the neighborhoods where they reside.
\end{itemize}

\section*{Data and Methodology}

\label{sec:methodology}
\subsection*{Nextdoor Data}
\label{sec:dataset_nd}
\pb{Nextdoor primer.} Nextdoor is a location-based social network with over 270K registered neighborhoods in 11 countries and over 10 million users~\cite{nextdoor-stats}. Nextdoor divides geographical areas into neighborhoods. Nextdoor assigns users to the \textit{neighborhood} where they  resides. To ensure that a user is a neighbor of a particular neighborhood, new users  validate their home addresses via regular (snail) mail. 

For each neighborhood, Nextdoor creates a dedicated forum where users can post and interact (\eg reply, and react to each other's posts).
Users exclusively interact with their neighbors, \ie the users of the neighborhood they are associated with. 
As a result, the data from a neighborhood exclusively includes the posts of the users that have validated their location in that geographical area.

In the rest of this paper, we use the term 
\textit{neighborhood} to refer to the specific areas into which Nextdoor divides a region and \textit{neighbor} to refer to the Nextdoor user registered as living in a neighborhood. 

\pb{Neighborhoods.}
We collect
2,201,051 posts from 64,283 USA neighborhoods and 351,894 posts from all the 3,325 neighborhoods in the 10 UK cities with the largest population. UK cities include London, Birmingham, Liverpool, Sheffield, Bristol, Glasgow, Edinburgh, Leeds, Manchester, and Bradford.    
Due to data scraping limitations, We have incomplete data for 13 USA states (see the percentage of neighborhoods for which we collected data in parenthesis): Texas (90.69\%), California (88.6\%), Georgia (80.1\% ), Florida (84.43\%), Alaska (91.17\%), Washington (78.6\%), Wisconsin (85.81\% ), Virginia (80.51\%), Alabama (82.62\%), Nevada (88.02\%), New Jersey (78.4\% ), Louisiana (87.23\%), and New Mexico (91.1\%).  
Overall our data includes 76.2\% of neighborhoods in the USA,  covering  65.8\% of the population in the USA. 
For the 10 UK cities, our data is complete, \ie the data collected covers all their neighborhoods.
Our data includes almost one full year (November 2020 -- September 2021). Note that we count Washington, District of Columbia (DC), as a state because of its autonomous status. 

We obtain a list of neighborhood names directly from the Nextdoor website for the USA~\cite{nextdoor-neighborhoods}.
For the UK, a list of neighborhoods does not exist on Nextdoor. Instead, we collect the list of UK neighborhoods by crawling the map integrated into Nextdoor. Due to the resource-intensive nature of this process, we collect data for the neighborhoods of the ten most populous UK cities.

\pb{Neighborhood location.}
For each neighborhood, we obtain neighborhood names and locations (\ie latitude and longitude). For the UK, we obtain the latitude and longitude of a neighborhood while crawling the Nextdoor map.
For the USA, we employ the GeoPy API to map the coordinates of a neighborhood to its zip code~\cite{geopy-python}. 
We then map the coordinates of each neighborhood into 
the lowest geographical granularity at which official statistical data is reported:
Zip code for USA, and
Lower Layer Super Output Area (LSOA) in the UK, a small geographic area used for UK's statistical reporting. 
To map a UK postcode to its LSOA, we use official data~\cite{lsoa-uk}.

\pb{Neighbors and posts.}
We then iterate through each neighborhood to collect posts written by its neighbors. 
In total, we obtain over 2.6M posts from 67,608 neighbors.
Table~\ref{tab:stats_data} shows the main attributes of the  data.

\begin{table}
\normalsize
\centering
\begin{adjustbox}{max width=\columnwidth}
\begin{tabular}{|c|c|c|c|}
\hline
\textbf{Attributes}& \textbf{USA}  & \textbf{UK}& \textbf{Total}   \\ \hline
Posts                        & 2,201,051       & 351,894 & 2,602,045        \\ \hline
Neighborhoods       & 64,283        & 3,325 & 67,608          \\ \hline
Cities                       & 5,849          & 10 &5,859            \\ \hline
zip code(USA)/LSOA(UK)      & 30872 & 2512          &33284  \\ \hline
Comments & 17,421,050 & 2,246,814 &19,667,864 \\ \hline
Neighbors     & 6,6480,730 & 1,744,948 & 68,225,678            \\ \hline
\end{tabular}
\end{adjustbox}
\caption{Nextdoor dataset.}
\label{tab:stats_data}
\end{table}

\subsection*{Feature Engineering}
\label{sec:data_aug}
\pb{Crime data, population, and income.}
For each neighborhood, we obtain socioeconomic data at the zip code and LSOA level by querying governmental databases of official statistics. 
For the USA zip codes, 
we obtain the population and median annual income from the latest Census \cite{census-usa}. 
We obtain crime per 10,000 people data from FBI's Crime Data Explorer \cite{fbi-usa}. 
For the UK LSOAs,
we collect the median annual income and population data from the UK's Office of National Statistics \cite{population-uk} using its latest Census update (2021) and crimes per 10,000 people from the UK Metropolitan Police \cite{met-uk}.

\pb{Inequality.}
Income inequality data is unavailable at the neighbor-level.
Instead, we calculate inequality at the neighborhood level: we measure the inequality between the group of neighborhoods that are in the same vicinity. 
As an inequality metric we compute the Atkinson Index for each neighborhood~\cite{atkinson1992economic}:

\begin{equation}
    A(y_{1},.....,y_{n})=
1-\frac{1}{\mu}(\frac{1}{N}\sum_{i=1}^{N}y_{i})
\end{equation}

\noindent where $A$ is the Atkinson index of neighborhood $i$, 
\textit{y$_i$} is its income, 
\textit{N} is the set of nearby neighborhoods (including $i$), 
$\mu$ is the mean income of the $N$ neighborhoods.
Note that the Atkinson index uses an
inequality aversion parameter which we equal to zero, to avoid any assumptions about the impact of inequality.
A set of $N$ neighborhoods with identical income will have an Atkinson Index equal to zero. The greater the difference between the incomes, the closer to 1 the index is.

We then compute each neighborhood's set of $N$ nearby neighborhoods.
To define the vicinity of a neighborhood, we identify the minimum radio that still renders at least another nearby neighborhood for each target neighborhood. 
We compute the distance between each pair of neighborhoods with the Haversine formula~\cite{inman1849navigation}. 
We find that a minimum radius of 24.92 and 2.97 miles for the USA and UK is necessary to have non-empty surroundings for all neighborhoods. 
We round up these numbers and use a radio of 25 and 3 miles for the USA and UK, respectively.  
We also discover 418 USA neighborhoods more than 100 miles from the nearest neighborhood. 
We verify that they are, in fact, in remote, isolated areas. Furthermore, due to the incomplete scraping of USA neighborhoods, many remote, isolated neighborhoods are not in our dataset.

\pb{Crime-related posts.}
To assess how sensitive to crime a neighborhood is,
we identify the posts discussing crime and compare the number with the actual crime reported  in that neighborhood.
We focus on crime because it is the only topic for which we found official statistics with rich and detailed geolocated data for both countries. Additionally, manual inspection revealed that Nextdoor users commonly discuss crime in their neighborhoods.

We label each post according to whether it discusses crime and the type of crime using the Semantic Search application of the pre-trained \textit{msmarco-distilbert-base-v4} Sentence-BERT (S-BERT) model~\cite{reimers-2019-sentence-bert}.
This model uses siamese and triplet network architectures to generate semantically significant sentence embedding. 
We then compute the cosine similarity between the embedding of a post and the description of the official crime category. 
We obtain the official categories of a crime from the FBI's Crime Data Explorer \cite{fbi-usa}, and the UK Metropolitan Police's Crime \cite{met-uk}. We note that the categories of the same type of crime vary between the two countries.
In particular, the USA definition of violent crimes is narrower than in the UK.
While this does not affect within-country comparisons, we caution about the USA--UK comparisons. For consistency, we only consider categories in both countries' statistics. Additionally, we group crime categories into  three major categories; (i) Drugs and Order, (ii) Theft and Property Damage, and (iii) Weapons and Violent Crimes. See the complete list of crime categories in Table \ref{tab:crime_categories}.

We consider that a post discusses a crime whenever the similarity is greater than a threshold.
To determine the threshold,
our native English-speaking human annotator takes a random sample of 2,000 posts, 1,000 from each country from S-BERT tagged data, and manually annotates them as crime or no-crime discussions. 
When a post discusses a crime, our annotator also annotates the type of crime discussed.
We repeat this process for several thresholds and use Cohen's Kappa score (K)~\cite{mchugh2012interrater} between the S-BERT and our manual annotation as the criteria for the threshold. 
Table~\ref{tab:threshold-K} reports the thresholds and the corresponding K.
We choose 0.7 as our threshold, which maximizes K for the USA and UK (K=0.971, K=0.944 in the USA and UK, respectively).
We then use this 0.7 threshold to label 543,459 USA posts and 124,763 UK posts as crime-discussing.

\begin{table}[h!]
\centering
\normalsize
\begin{adjustbox}{max width=0.65\columnwidth}
\begin{tabular}{|l|l|l|} 
\hline
\textbf{Cosine Similarity} & \textbf{K (USA)} & \textbf{K (UK)}  \\ 
\hline
0.5                                  & 0.875                                  & 0.847                                  \\ 
\hline
0.6                                  & 0.948                                  & 0.895                                  \\ 
\hline
0.7                                  & 0.971                                  & 0.944                                  \\ 
\hline
0.8                                  & 0.939                                  & 0.917                                  \\ 
\hline
0.9                                  & 0.926                                  & 0.891                                  \\
\hline
\end{tabular}
\end{adjustbox}
\caption{Cosine similarity thresholds and their respective Cohen's Kappa Score (K).}
\label{tab:threshold-K}
\end{table}
\begin{table*}[!pbt]
\normalsize
\centering
\begin{adjustbox}{max width=\linewidth}
\begin{tabular}{|l|l|l|l|l|l|l|l|l|} 
\hline
                       & \multicolumn{2}{c|}{\textbf{Population}} & \multicolumn{2}{c|}{\textbf{Posts}} & \multicolumn{2}{c|}{\textbf{Neighborhoods }} & \multicolumn{2}{c|}{\textbf{Neighbors }}  \\ 
\hline
                       & USA  & UK                                & USA  & UK                           & USA  & UK                                    & USA  & UK                                 \\ 
\hline
\textbf{Population}    & 1    & 1                                 & 0.83 & 0.96                         & 0.93 & 0.96                                  & 0.96 & 0.97                               \\ 
\hline
\textbf{Posts}         & 0.83 & 0.96                              & 1    & 1                            & 0.91 & 0.97                                  & 0.84 & 0.89                               \\ 
\hline
\textbf{Neighborhoods} & 0.93 & 0.96                              & 0.91 & 0.97                         & 1    & 1                                     & 0.93 & 0.95                               \\ 
\hline
\textbf{Neighbors}     & 0.96 & 0.97                              & 0.84 & 0.89                         & 0.93 & 0.95                                  & 1    & 1                                  \\
\hline
\end{tabular}
\end{adjustbox}
\caption{Correlation between posts, population, neighborhoods, neighbors, and official population.}
\label{tab:corr_usa_uk_posts}
\end{table*}

\pb{Post sentiment.} To assess how positive a neighborhood's posts is, we label each post's sentiment in our dataset
with a pre-trained Valence Aware Dictionary and Sentiment Reasoner (VADER) model~\cite{hutto2014vader}.
VADER is a lexicon and rule-based sentiment analysis instrument that outperforms the typical human reader in its sensitivity to assumptions transmitted in web-based media (VADER's F1=0.96, Human F1=0.84) \cite{hutto2014vader}. VADER works well on content-aware text with emotions, such as social media text. As people use Nextdoor as a social network and Nextdoor itself encourages people to post using few words, VADER is appropriate for Nextdoor data.

\pb{Text embedding.}
To investigate whether the text posted generally differs between richer and poorer neighborhoods, 
we obtain semantic features of the posts via embedding. 
We preprocess every post (removing mentions, URLs, \etc) 
and discard neighborhoods with less than 10 posts. 
Our final dataset consists of 63,587 neighborhoods from USA and 3,282 neighborhoods from the UK.
Then, we convert each post's text into a sentence embedding using
the best-performing pre-trained sentence transformer model, ``all-mpnet-base-v2'' powered by \textit{Hugging-Face} \cite{huggingface}.
This model is tuned to map every sentence or short paragraph to a 768-dimensional vector space while preserving important text features. 
We then aggregate the sentence embedding of all the posts of a neighborhood.
Similarly to~\cite{Arora2017},
we employ an element-wise mean-pooling aggregation method.

We obtain a 768 dimensions (mean-pooled) feature per neighborhood. 
This large number is problematic due to the
\one Curse of Dimensionality effect~\cite{Aggarwal2001},
and \two~some features would outweigh the effect of others if we were to simply concatenate them. 
We avoid these problems by reducing the textual embedding with Uniform Manifold Approximation and Projection (UMAP)~\cite{McInnes2018}, the state-of-the-art method.
We reduce the 768 dimensions to 5, as suggested in~\cite{zhang2022neural, jimenez2021uncovering}.

\subsection*{Data Representativity}
We observe that the 20 most populated USA states (40\%) account for 70\% of the posts, 72\% of the neighborhoods, and 69\% of the neighbors.
The UK is more skewed, with London concentrating 70\%, 61\%, and 76\% of the UK's posts, neighborhoods, and neighbors, respectively. 

\pb{Methodology.}
To assess whether this reflects the residents' concentration in those areas, we look at the correlation between Nextdoor data attributes and the official population distribution in Table~\ref{tab:corr_usa_uk_posts} using Pearson correlation coefficient \cite{lee1988thirteen}.
We calculate the correlation at the neighborhood level (\eg users in a neighborhood and the corresponding official population), except for neighborhoods, which we do at the city level.
We find a high correlation for each attribute, with a maximum of 0.97
(USA neighbors to population) and a minimum of 0.83 (USA posts to population). These high numbers give us confidence that the data of Nextdoor is representative of the USA and UK populations.
See the Appendix for more details on the distribution of Nextdoor activity and population across USA states and UK cities.

\begin{figure}[!b]
	\begin{center}
		\includegraphics[width=\columnwidth]{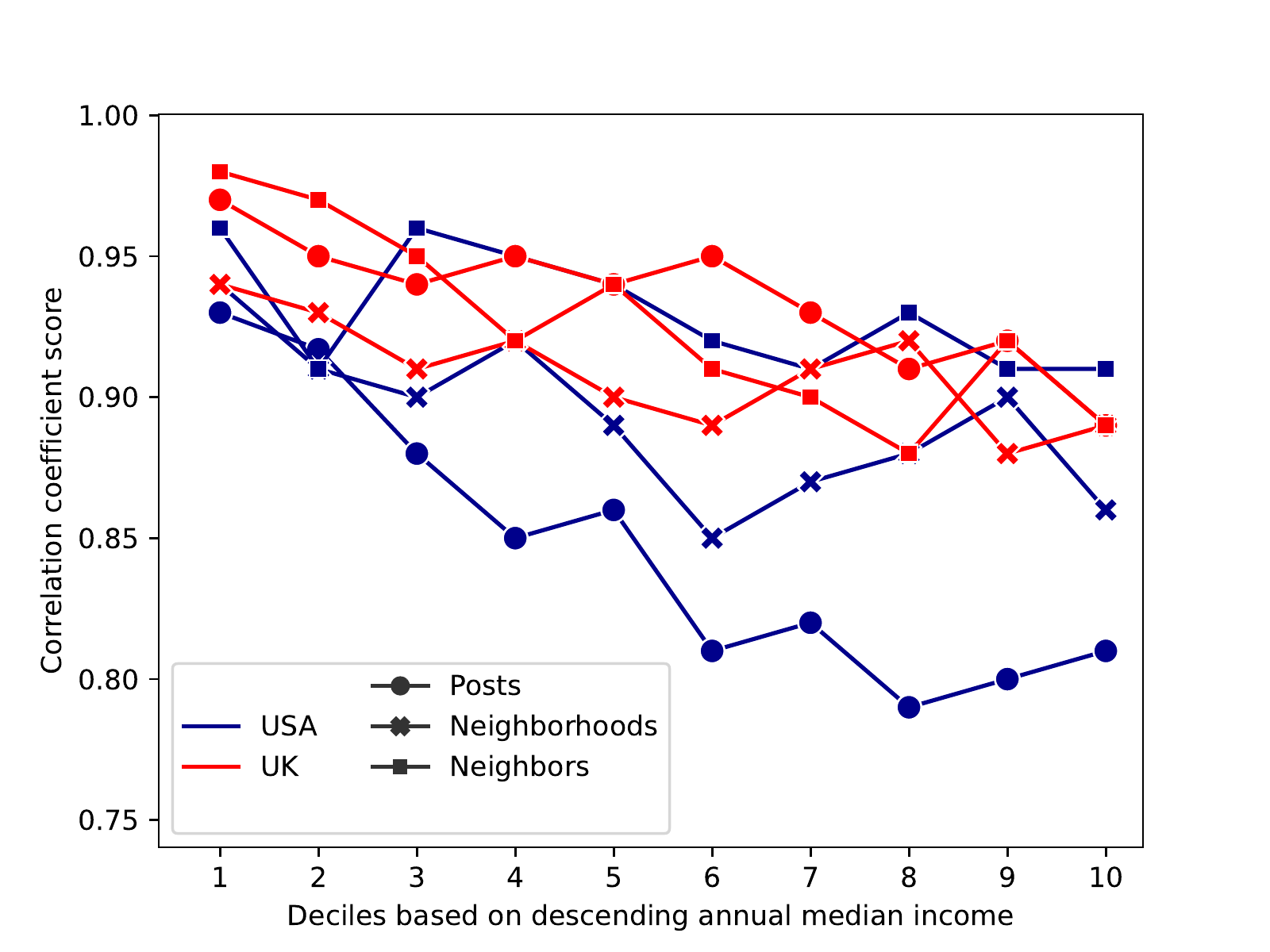}
	\end{center}
	\caption{Correlation between posts, neighborhoods, neighbors, and official population over income deciles (from richer to poorer).}
	\label{fig:corr_feature}
\end{figure}

\pb{Income and representativity.}
To examine how well-represented neighborhoods of different income levels  are, we look at the correlation between Nextdoor attributes and the underlying populations in Figure~\ref{fig:corr_feature} (similarly to Table~\ref{tab:corr_usa_uk_posts}). We find a remarkably high correlation between the number of posts, neighbors, neighborhoods, and the underlying population for both the USA and the UK.
Correlations range between 0.88--0.98 for the UK and 0.79--0.93 for the USA, with most correlations above 0.9.
These correlations tend to be higher for richer deciles and are usually higher for the UK.
Further investigation shows that in the USA, residents of rich neighborhoods are more likely to be registered in Nextdoor. 
We find that the USA ratio of population-to-users (at the neighborhood level) tends to be indeed higher for the richest deciles (below 8) than for the poorer ones (below 18). See Figure \ref{fig:nextdoor-one-official} in Appendix for more details.

Throughout the rest of the paper,
we compare the richest 20\% and the poorest 80\%.
Since the richest deciles are better represented and the rest are aggregated together, this will help ensure that our results remain representative of those levels of income.
For the USA, 29.87\% neighborhoods fall in the richest 20\%, and 70.13\% are part of the poorest 80\%. 
In the UK,
42.5\% neighborhoods are in the richest 20\%, and 57.5\% are in the poorest 80\%.
We again caution regarding cross-country comparisons due to these differences.

\section*{Income and Online Discussions}
\label{sec:income_diff}

With the above data, we now examine whether income differences result in different online behaviors. First, we study differences in crime discussions between richer and poorer neighborhoods. 
Then we compare the sentiment of  posts for the same income categories. 

\subsection*{Income and Crime Discussions}
\label{subsec:income_crime}
\begin{figure}[!b]
	\begin{center}
    \includegraphics[width=\columnwidth]{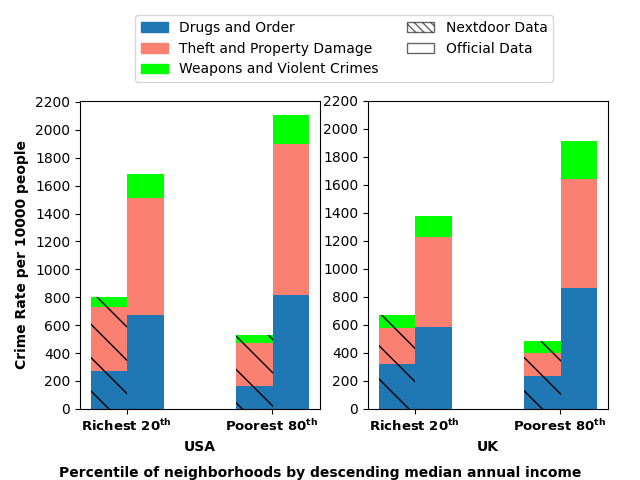}
	\end{center}
	\caption{Official crime and crime discussion rates (per 10,000 people) in Nextdoor for  the richest 20$^{th}$ and poorest 80$^{th}$ percentile neighborhoods.}
	\label{fig:crime_all}
\end{figure}

\pb{Richer neighborhoods seem to be more sensitive about crime.} Figure~\ref{fig:crime_all} shows the official crime rate per 10,000  people 
and the crime-discussion rate per 10,000 Nextdoor neighbors for the 20\% richest and 80\% poorest neighborhoods in USA and UK.
We find that
neighbors in rich neighborhoods discuss crime more often than those in poorer ones.
Interestingly, this is the case even though the actual crime rate is substantially higher in poorer neighborhoods. To further analyze this,
we breakdown crimes into three types using the official crime categories: 
\one~Drugs and Order;
\two~Theft and Property Damage;
and
\three~Weapons and Violent Crimes. 
See more details about considered crime categories in Table \ref{tab:crime_categories}.

\pb{Non-violent crimes are discussed more than violent crimes on Nextdoor.} 
We observe that non-violent crimes are discussed more than weapons and violent crimes across all neighborhoods.
\begin{figure}[t]
	\begin{center}
    \includegraphics[width=\columnwidth]{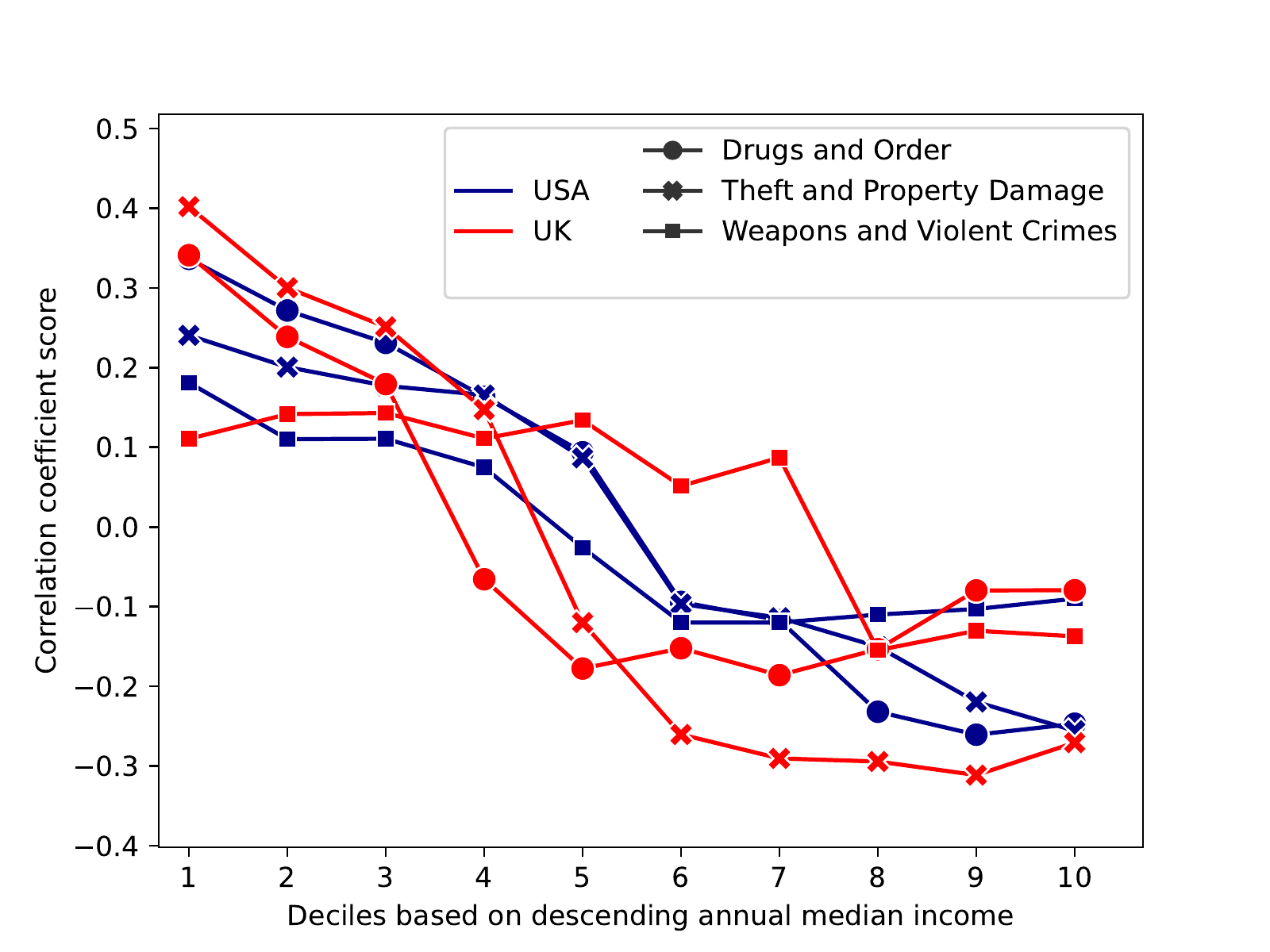}
	\end{center}
	\caption{Correlation between crimes per 10000 people
reported officially and discussed on Nextdoor over income deciles (from richer to poorer).}
	\label{fig:crime_corr}
\end{figure}
To analyze these trends, we calculate the Pearson correlation coefficient between the rates of official crime rate and Nextdoor crime discussions. 
Figure~\ref{fig:crime_corr} shows these correlations across crime types for the USA and UK from richest (left) to poorest percentiles (right). We observe a clear trend across all types of crime and countries: richer neighborhoods discuss these crimes proportionally to their actual occurrence. This changes abruptly as we move towards poorer neighborhoods,
with the poorest neighborhoods discussing crime disproportionately to the actual crime occurrence.

Whereas most trends are very similar for both USA and UK, weapons and violent crimes show different behavior. These crimes are discussed proportionally more in the USA than in the UK for the richer neighborhoods, but the middle-income neighborhoods in UK tend to discuss this type of crime more than their USA counterparts. As we mentioned in Data and Methodology Section, 
the definitions of crime in the USA and the UK differ. This is particularly true for violent crime and possibly responsible for these differences.

\subsection*{Income and User Sentiment}
\label{sec:income_sentiment}
\begin{figure}[t]
	\begin{center}
    \includegraphics[width=\columnwidth]{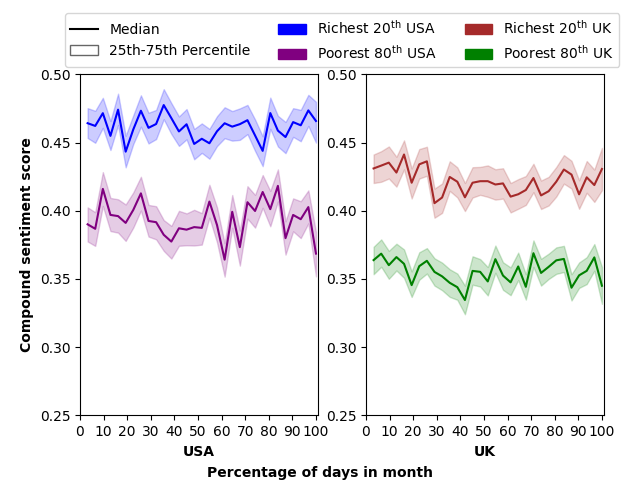}
	\end{center}
	\caption{Median compound sentiment score for the richest 20\% and poorest 80\% neighborhoods.}
	\label{fig:sentiment_box}
\end{figure}

As prior work indicates that income and self-reported positive sentiment are usually correlated~\cite{easterlin1974does,giorgi2021well, cui2022social}, we now look at the relationship between income and post sentiment. We calculate the median of the sentiment scores for 
the posts of the neighborhoods in the same income group and each \% of the duration of the month. Figure~\ref{fig:sentiment_box} shows the median compound sentiment scores across all months in our dataset for the 20\% richest and 80\% poorest neighborhoods.
Note that the higher the sentiment the more positive the text is. 

\pb{Richer neighborhoods seem to be more positive.} Our analyses show clear differences in user post sentiment across income levels. Posts from richer neighborhoods have more positive sentiment both in the USA and UK, where the richest 20\% have a clearly higher sentiment than the bottom 80\%. 

\subsection*{Income and Text Features}
\label{sec:income_text}
The previous analysis on crime discussion and post sentiment points to clear differences in online discourse depending on the income of a neighborhood. We now examine whether the text of the posts, in general, also reveals these differences.
We employ the text embedding and dimensionality reduction described in Data and Methodology Section 
to study whether the content differs depending on the income of a neighborhood.
\begin{figure}[!b]
	\centering
	\begin{subfigure}[b]{0.48\columnwidth}
		\centering
		\includegraphics[width=\linewidth]{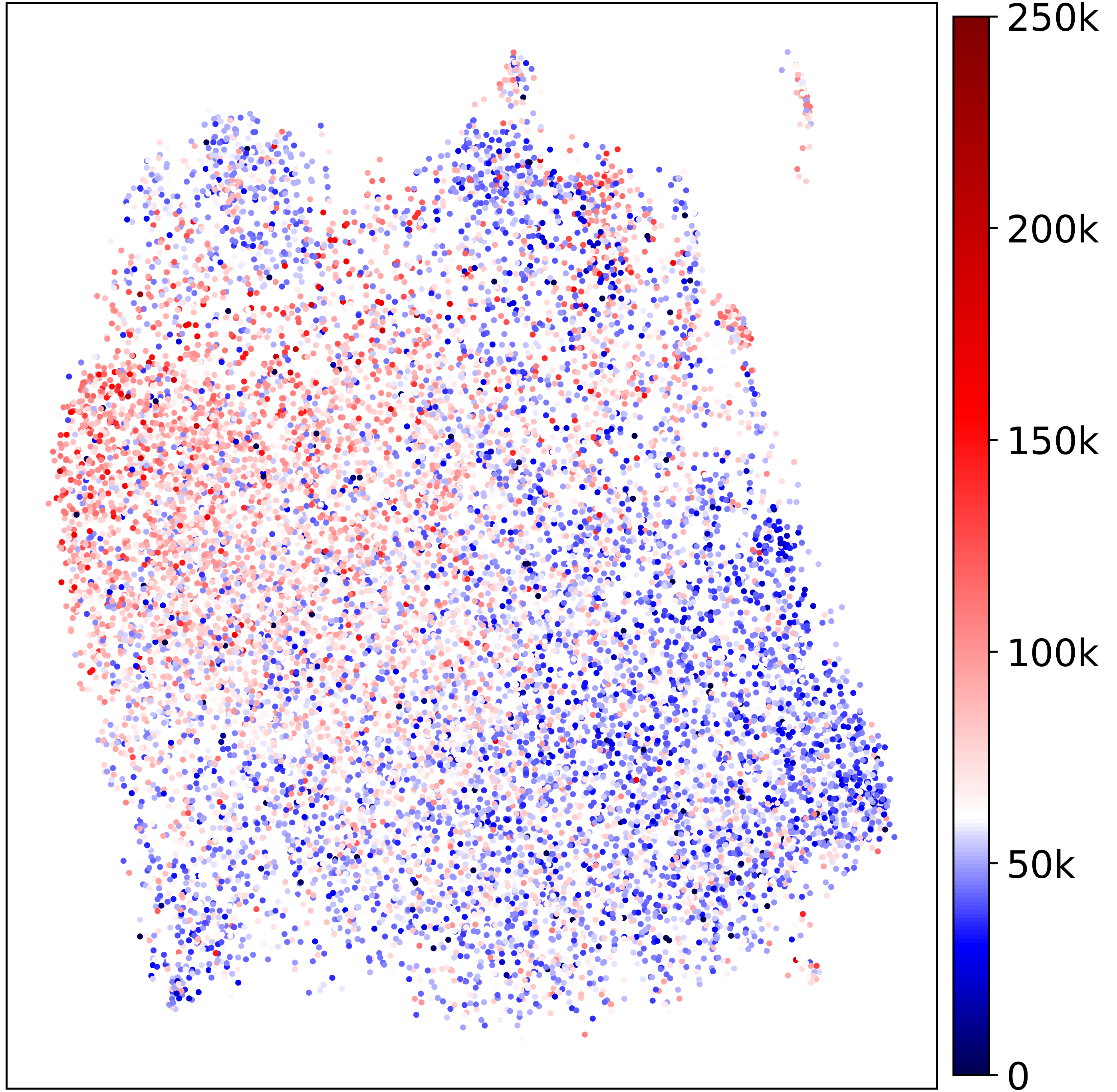}
		\caption{US}
		\label{fig:us_income}
	\end{subfigure}
	\hfill
	\begin{subfigure}[b]{0.48\columnwidth}
		\centering
		\includegraphics[width=\linewidth]{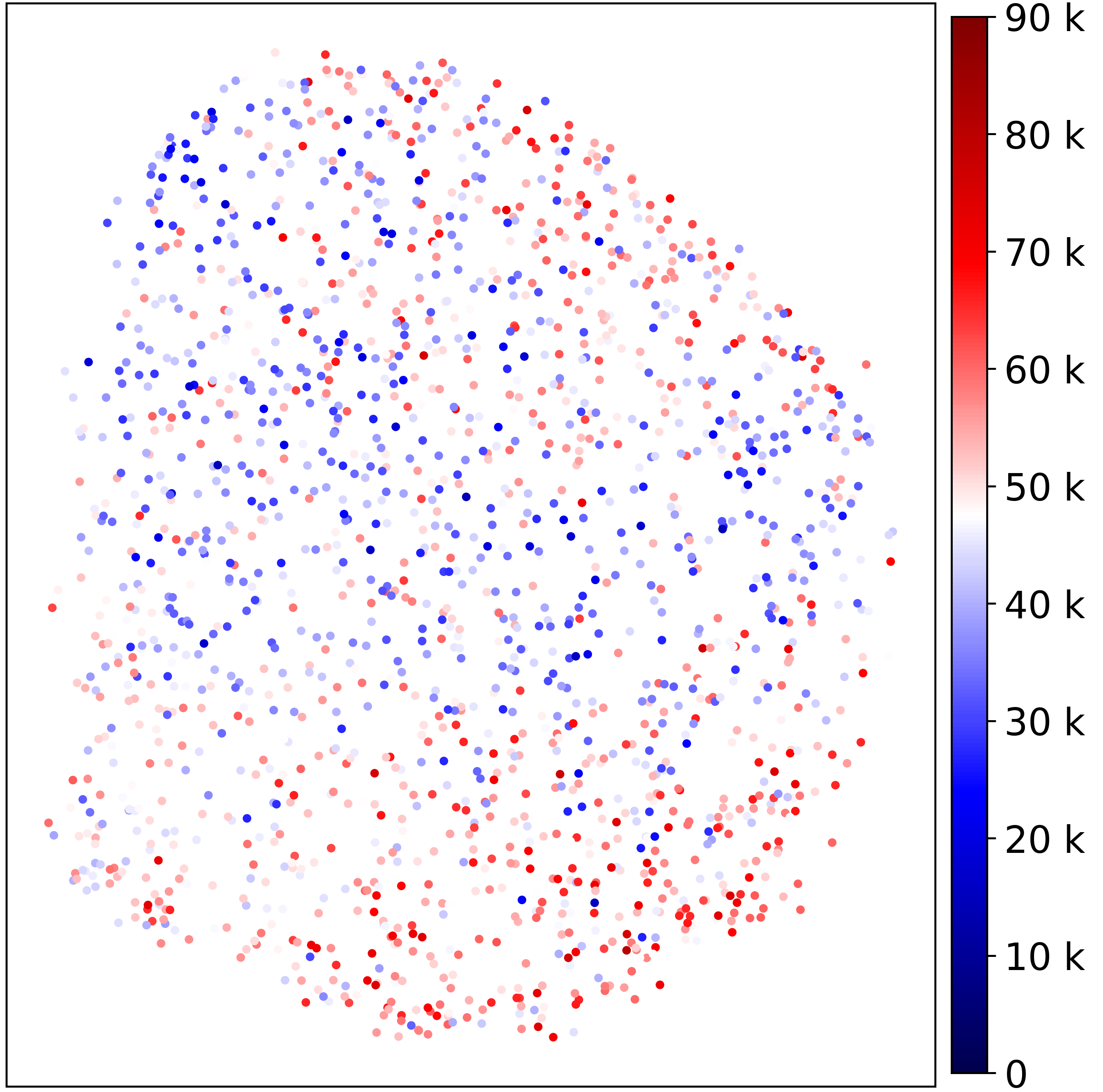}
		\caption{UK}
		\label{fig:uk_income}
	\end{subfigure}
	\hfill
		\caption{2D projection of semantic features posted from every neighborhood. The ``seismic'' colormap represents the lowest-income neighborhood with the darkest blue and the highest-income with the darkest red.}
\label{fig:embed_income}
\end{figure}
Figures~\ref{fig:us_income} and \ref{fig:uk_income} depict the 2D projection of the embedding vectors per location in the USA  and UK, respectively.  

The dots, which represent locations, are color-coded in a ``seismic'' continuous colormap to portray the highest-income locations as dark red, and the lowest-income neighborhoods as dark blue. 
The closer the income is to the median income, the lighter the color is. The level of separation between the two colors indicates the level of online content polarization based on income. 
Since blue and red dots are visibly separated from each other in Figure~\ref{fig:embed_income}, we confirm that the text posted by users in poor neighborhoods  is semantically distinguishable from the texts generated in wealthier neighborhoods.

The income-based semantic differences in the user-generated text can stem from various qualities such as discussed topics. We showed that this is the case for income, where we had detailed and geolocated official data, but it could be true for others themes as well. 
As the semantic encoding provided by sentence transformers is black-box in its nature, understanding the underlying aspects of these semantic differences is beyond the scope of this paper.

\section*{Income Inequality and Online Discussions}
\label{sec:income_inequality}
In previous sections, We demonstrate how income manifests in online discussions. In this section, we explore whether income inequality also plays a role.
Unfortunately, there is no official income inequality data available at the neighborhood level. 
Instead, we calculate how equal or unequal a neighborhood and its surroundings are.
For each neighborhood, we identify other neighborhoods within its vicinity as described in Data and Methodology Section.
We then calculate the Atkinson score for the set of neighborhoods across the surrounding area (including the target neighborhood). This gives us a metric of how unequal the vicinity is at the neighborhood level, \ie it does not reveal how unequal neighborhoods are within themselves.

To see how income and inequality (as defined above) interact, we classify neighborhoods depending on their income and the inequality level of their surroundings.
We again classify neighborhoods into the 20\% richest and the 80\% poorest.
We also classify the vicinity of a neighborhood into the 20\% most equal and the 80\% most unequal.
When combining income and inequality classifications, we identify 4 categories of neighborhoods:

\begin{itemize}
    \item \textbf{RE}: Richest neighborhoods and most Equal vicinity.
    \item \textbf{RU}: Richest neighborhoods and most Unequal vicinity.
     \item \textbf{PE}: Poorest neighborhoods and most Equal vicinity.
    \item \textbf{PU}: Poorest neighborhoods and most Unequal vicinity.
   
\end{itemize}

\subsection*{Income Inequality and Crime Discussion} 
\label{subsec:inequality_crime}

\begin{figure}[!h]
	\begin{center}
    \includegraphics[width=0.95\columnwidth]{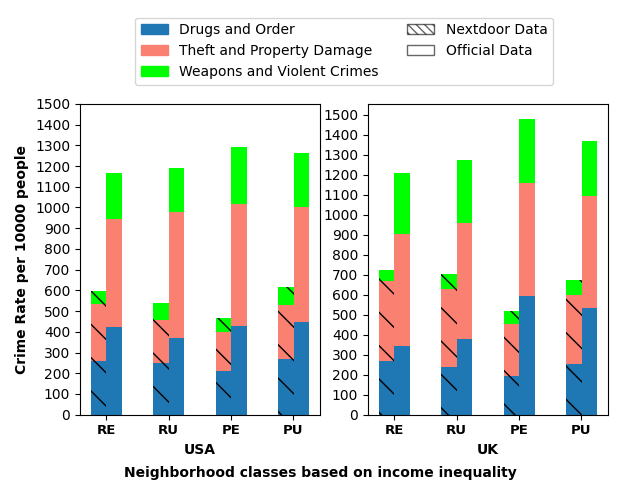}
	\end{center}
	\caption{Crime discussion rate (Nextdoor)  and official crime rate per 10,000 people.}
	\label{fig:crime_inequality}
\end{figure}

We investigate how crime is discussed in neighborhoods according to their income and the income inequality with their vicinity. We again find remarkably similar trends for the USA and UK. Figure~\ref{fig:crime_inequality} 
shows the crime discussed in Nextdoor and officially reported per 10,000 people for the 4 categories of income-inequality.

We observe that the Richest and Equal neighborhoods (RE), seem to be more crime-sensitive as crime is discussed there more often. While the difference is small (particularly in the UK), since the official rate is higher for the neighborhoods in more unequal vicinities (RU), residents indeed discuss crime proportionally more. 
\begin{figure}[!b]
	\begin{center}
    \includegraphics[width=0.95\columnwidth]{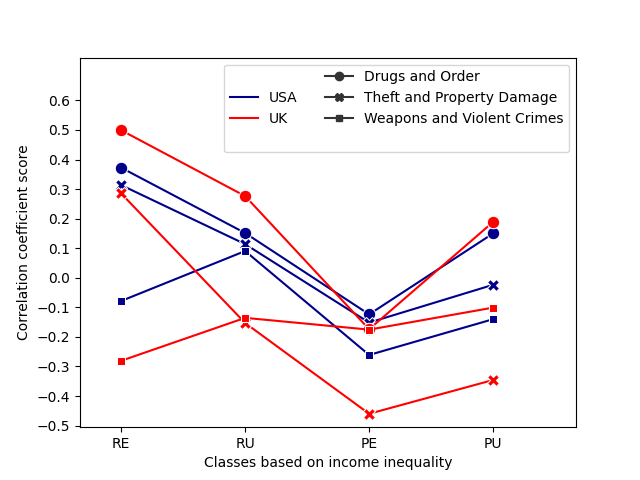}
	\end{center}
	\caption{Correlation between crimes (per 10,000 people) discussed (Nextdoor) and officially reported in neighborhoods according to their income and inequality.}
	\label{fig:corr_inequality}
\end{figure}

Exactly the opposite is true for the poorest neighborhoods.
For the poorest neighborhoods,
crime is less discussed when the income of vicinity is more equal (PE) than when it is more unequal (PU). 
This is true even though the actual crime goes in the opposite direction:
higher for more income-equal vicinities than for unequal ones  in poorer neighborhoods.
This indicates that the higher the income within a neighborhood or its surroundings, the more it is likely for crime to be discussed online.

We then calculate the Pearson correlation coefficient between the official crime rate and crime discussed in Nextdoor for each category in Figure~\ref{fig:corr_inequality}.
This figure confirms previously observed trends between crime rate and crime discussion for all categories except for ``weapons and violent crimes'' which follow the opposite trend in the richest neighborhoods, 
\ie most equal neighborhoods discuss this type of crime proportionally less than the more unequal ones.

\subsection*{Income Inequality and User Sentiment}
\label{subsec:inequality_sentiment}
We now look at the relationship between income inequality and sentiment at the neighborhood level. 
We calculate the median of the sentiment scores for the posts of the neighborhoods in the same income inequality classes and each \% of the duration
of the month. 

\begin{figure}[t]
	\begin{center}
    \includegraphics[width=\columnwidth]{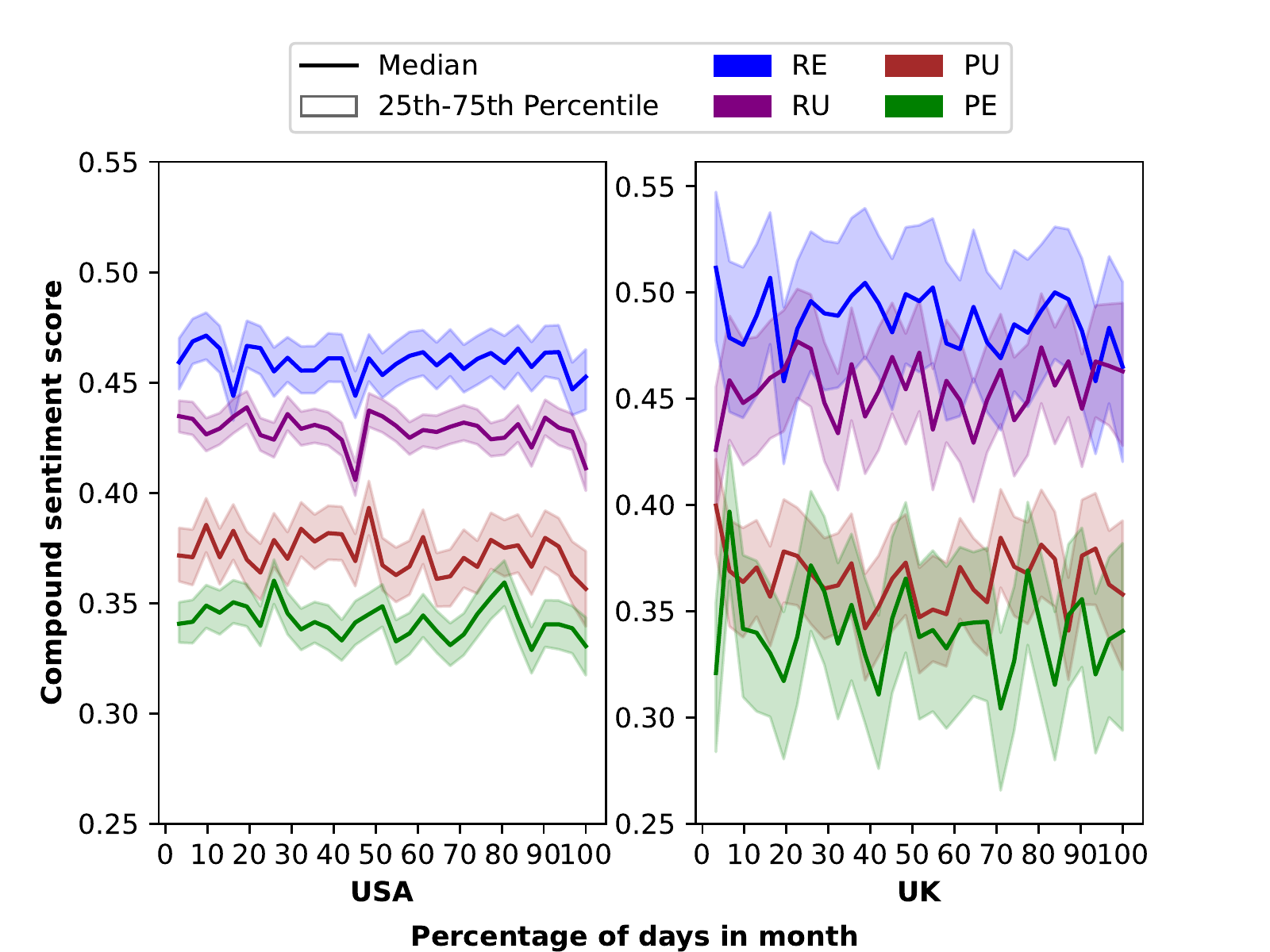}
	\end{center}
	\caption{Median compound sentiment score of Nextdoor posts in neighborhoods with different income and inequality.}
	\label{fig:sentiment_inequality}
\end{figure}

Figure~\ref{fig:sentiment_inequality} shows the median compound sentiment scores across all days of a month (in percentage) in our dataset for all income inequality classes. 
Note that the higher the sentiment the more positive the text is. 

Richer neighborhoods again seem more positive:
the richest neighborhoods tend to have a higher positive sentiment regardless of the level of inequality in their vicinity. 
Richer neighborhoods with more equal vicinities (RE) have a more positive sentiment. We observe the opposite in the poorest neighborhoods, 
more equal vicinities are associated with a lower sentiment (PE).
These findings are again almost identical for both countries and again indicate that the higher the income within a neighborhood or its surroundings, the more positive the sentiment of its posts.

\section*{Evaluation and Validation}
\label{sec:income_prediction}
The previous sections demonstrate that neighborhoods with different income profiles exhibit clearly different textual traits. 
We now asses whether these differences are enough to predict whether posts are from a richer or poorer neighborhood.
We show that commonly used models succeed in this prediction task.

\pb{Text can predict income differences.}
After splitting data into 70\% train and 30\% test sets, we use a Lasso regression to confirm that the text embeddings used in the Income and Text Features Section 
indeed differ depending on the income level.
Table~\ref{tab:semantic_r2} shows the corresponding coefficient of determination (R$^2$) demonstrating that the text has a strong predictive power of the income of a neighborhood. 

\begin{table}[!tpbh]
\centering
\small
\begin{tabular}{|l|l|l|l|l|l|l|l|}
\hline
\textbf{Country} & \textbf{Train R$^2$} & \textbf{Test R$^2$}\\
\hline
USA &  0.39 & 0.35\\
\hline
UK &  0.69 & 0.59\\
\hline
\end{tabular}
\caption{Determination coefficient (R$^2$) of a Lasso regression exclusively using sentence embeddings to predict the income of a neighborhood.}
\label{tab:semantic_r2}
\end{table}

\begin{table*}[!tbph]
\centering
\small
\begin{adjustbox}{width=0.95\linewidth}
\begin{tabular}{|l|l|l|l|l|} 
\hline
\multicolumn{1}{|c|}{\textbf{Feature}}                    & \multicolumn{4}{c|}{\textbf{Feature Importance}}                                                                                                                                             \\ 
\hline
& \multicolumn{1}{c|}{\textbf{Income}} & \multicolumn{1}{c|}{\textbf{Inequality}} & \multicolumn{1}{c|}{\textbf{Income, no inequality}} & \multicolumn{1}{c|}{\textbf{Inequality, no income}}  \\ 
\hline
Discussed-to-official crime (theft property damage) ratio & 0.162                                & 0.141                                    & 0.22                                                & 0.251                                                \\ 
\hline
Discussed-to-official crime (drugs public order) ratio    & 0.138                                & 0.125                                    & 0.181                                               & 0.192                                                \\ 
\hline
Atkinson index                                            & 0.124                                & Target                                   & N/A                                                 & Target                                               \\ 
\hline
Median annual income                                      & Target                               & 0.120                                    & Target                                              & N/A                                                  \\ 
\hline
Mean-pooled (5 dimensions) sentence embeddings of posts   & 0.119                                & 0.118                                    & 0.169                                               & 0.161                                                \\ 
\hline
Registered Nextdoor users             & 0.112                                & 0.093                                    & 0.143                                               & 0.136                                                \\ 
\hline
Median readability score of posts                         & 0.092                                & 0.079                                    & 0.102                                               & 0.098                                                \\ 
\hline
Median post length                                        & 0.078                                & 0.061                                    & 0.098                                               & 0.091                                                \\ 
\hline
Discussed-to-official crime (weapons violence) ratio      & 0.069                                & 0.083                                    & 0.087                                               & 0.071                                                \\
\hline
\end{tabular}
\end{adjustbox}
\caption{Feature Importance (FI) of each feature in the Random Forest model to predict neighborhoods' income and income inequality with different feature sets. Note that ``Target'' refers to the explained variable.}
\label{tab:income_features}
\end{table*}
\pb{Prediction features.}
Encouraged by these findings, we develop multiple features based on the analysis of the previous sections and investigate their predictive power. We first use UMAP for dimension reduction as prior work shows improvements in the performance~\cite{zhang2022neural, jimenez2021uncovering}:
we reduce 768-dimensional mean-pooled textual embeddings to 5 dimensions.
For feature selection, we employ a Random Forest (RF) as this is 
the best-performing method~\cite{chen2020selecting}.
We compute the relative importance of each feature using the Gini index~\cite{breiman2017classification}.
Table~\ref{tab:income_features} presents the relative importance of each feature in our prediction tasks.

We observe that crime discussions are highly correlated with both income and inequality.
This is particularly true for the non-violent crimes categories, which rank top in feature importance.
We notice that the importance of these variables increases when we remove inequality from the features list while predicting income and vice versa.

\begin{figure}[!b]
	\begin{center}
    \includegraphics[max width=0.9\columnwidth]{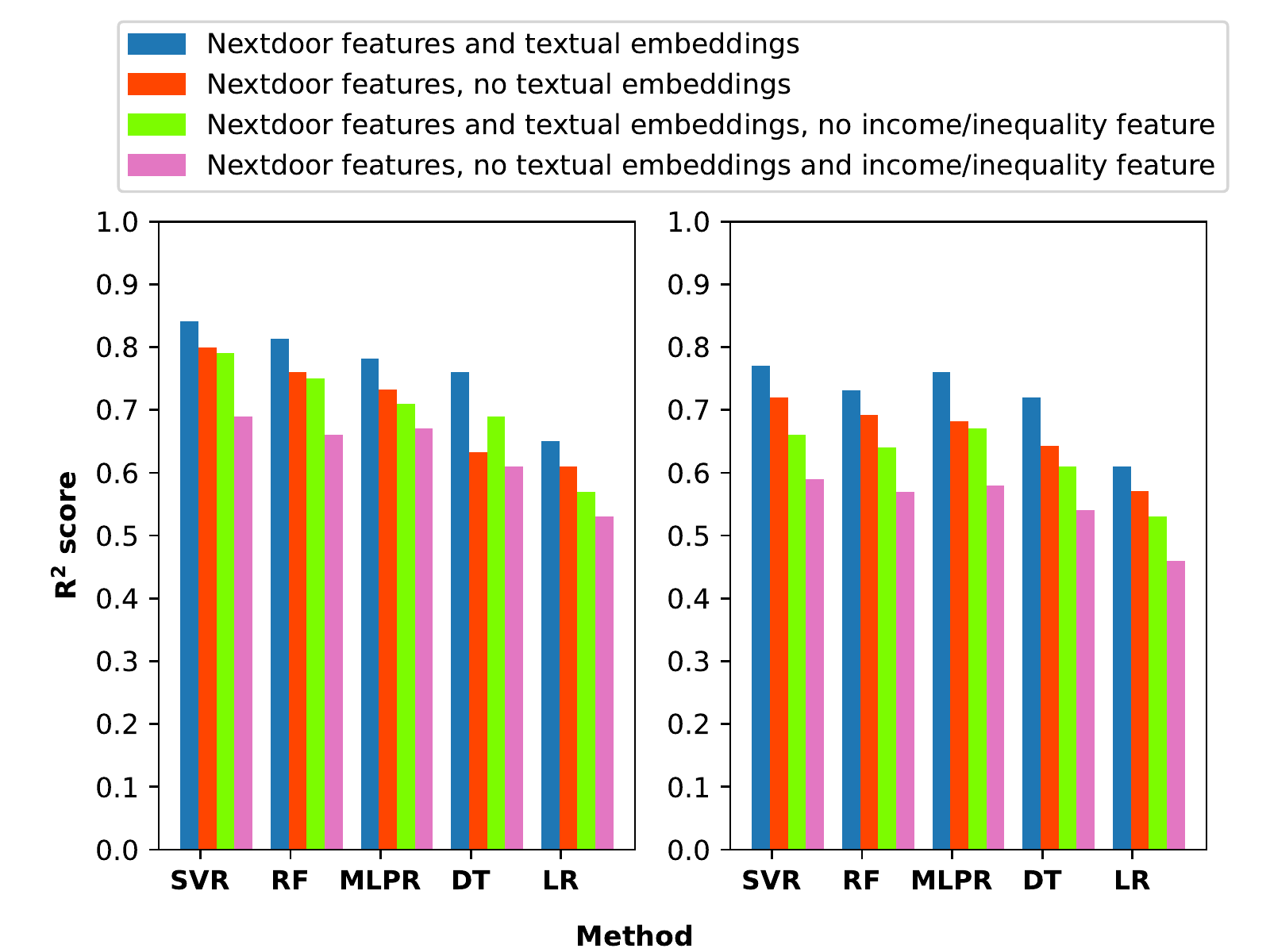}
	\end{center}
	\caption{Determination coefficient (R\textsuperscript{2})  for income and inequality predictions with Support Vector Regressor (SVR), Random Forest (RF), Multi-layer Perceptron Regressor (MLPR), Decision Trees (DT), and Linear Regression (LR).} 
	\label{fig:rsquared}
\end{figure}

\pb{Prediction Model Setup.} 
Again we split our data into 70\% train dataset and 30\% test dataset for our models.
We then train the regression models and use stratified 10-fold cross-validation as the evaluation approach. We implement Linear Regression (LR), Random Forest (RF), Support Vector Regressors (SVR), Multi-Layer Perceptrons-based Regressors (MLPR), Decision Trees, and Linear Regression (LR) models.
We define two tasks: 
\one predicting the annual median income of a neighborhood, and
\two predicting the inequality (\ie Atkinson index) of the vicinity of a neighborhood.

For each task, we train and test our models with and without text features and with and without inequality (when predicting income) and income (when predicting inequality).
We use the GridSearchCV method to optimize hyper-parameters for each algorithm. 
Details of our best model parameters chosen by the GridSearchCV method are presented in Table~\ref{tab:model_parameters} in the Appendix.

\pb{Model Evaluation.}
Our results show that the text posted online indeed reveals what is the income level and inequality of the area where they are posted.
Figure~\ref{fig:rsquared} summarizes the performance of our models. The prediction of income and inequality with a combination of textual and other Nextdoor features (income best R$^2$: 0.841, inequality best R$^2$: 0.77) has the highest performance.
We experiment training with fewer features and observe that even though the performance degrades, the R\textsuperscript{2} still high goes up to  0.686 for income and  0.59 for inequality with the most restrictive set of features (all Nextdoor features, but no textual embeddings and income/inequality feature). The predicting power for inequality is slightly lower than for income (between 7 and 13\% lower in the SVR model). 
This is natural as this prediction is inherently more difficult: 
a neighborhood in a highly unequal surrounding could be 
either a richer neighborhood surrounded by poorer ones, 
or the opposite, a poorer one surrounded by rich neighborhoods.

We are confident that our model is not overfitting our training data
given the low Test Root Mean-Square Errors of 0.159 and 0.429 for income and inequality in the best-performing model (see Figures~\ref{fig:rmse} and \ref{fig:rmse_inequality} in the Appendix).

\section*{Related Work}
\label{sec:relatedwork}
Much work has analyzed social networks from centralized social networking platforms such as Twitter and Reddit~\cite{iqbal2022exploring}
to the increasingly popular Mastodon~\cite{zia2023flocking} and other Fediverse platforms~\cite{bin2022toxicity}.

Previous research strove to infer the income of online users \cite{preoctiuc2015analysis, aletras2018predicting, lampos2016inferring, abitbol2018location}. For instance, 
for users that mention their profession in Twitter,
Preoctiuc et al. \cite{preoctiuc2015studying, aletras2018predicting} approximate the income of those users with the median income of such profession.
Then they predict the inferred user income with text features from their tweets (0.633 Pearson correlation). 
Differently, this work does not require heuristics: we use official statistics on the median income of the neighborhood where users reside.  
Moreover, our comprehensive data across USA and UK minimises possible sampling biases and successfully predicts neighborhood income  ($R^2=0.841$).

Similarly to our work, research using mobility and location-based application data can map individuals to a neighborhood and its official median income~\cite{moro2021mobility, toth2021inequality, xu2019quantifying}. 
Differently, this implies first inferring the residence based on the mobility patterns of a user. 
Additionally, there is limited online discourse that we could leverage for the purpose of this paper in mobility data and location-based apps such as Foursquare.

\section*{Discussion: Limitations and Future Work}
\pb{Bias.} 
There is bias towards richer neighborhoods and official data only reports the median income of a neighborhood. Since we are unaware of the distribution of income within a neighboorhood, we cannot fully asses the magnitude of the bias: Nextdoor users might tend to be the richest within their neighborhoods. Therefore our results could refer to some extent to  rich individuals living in neighborhoods of different incomes rather than to users with different income levels.

\pb{Data completeness.} 
We aim to have a complete dataset for our future analysis. Currently, our data is  incomplete for the USA (in 13 estates) and exclusively covers the 10 largest UK cities. This will allow us, to compare rural versus urban neighborhoods, for instance. 

\pb{Topics.} In our future work, we will analyze topics beyond crime. In this paper, we focused on crime because official statistics are rich and include the location of where a crime took place. Going forward we will study whether topics discussed differ across neighborhoods of different income levels. We also plan to focus again on topics for which there is geolocated data. In particular, we aim to look at how and whether politics manifest in the text posted by Nextdoor users and how politicians relate to the neighborhood they represent.

\pb{Income prediction.} Finally, we want to explore further income prediction. Our future work will explore whether a general model can predict the income of individuals from their online discourse regardless of the platform where it is posted. 
We will experiment with models trained with Nextdoor data and predict the income level of  users from other platforms (\eg Twitter).

\section*{Conclusion}
This paper provided the first large-scale analysis of Nextdoor. 
We showed that the online content generated by users reveals economic factors, including income and inequality.
We collected 2.6 Million posts from 64,283 neighborhoods in the United States (USA) and 3,325 neighborhoods in the United Kingdom (UK) from an un-studied online social network, Nextdoor.

A unique feature of our Nextdoor dataset is that users and posts are associated to the neighborhood where they reside.
We then linked the neighborhood of the users with official socioeconomic (\eg income, and crime) and showed that neighborhoods from different income levels post very different content.
We found that the richest neighborhoods seem more sensitive about crime: crime is discussed more than in the poorest neighborhoods, even though the actual crime rates are higher in the latter.
We also showed that the text posted by the richest neighborhoods seems to be more positive, with a consistently higher sentiment.
We found that income inequality is also visible  in online discourse and interacts with the income of the neighborhood.
The richest neighborhoods with the most equal surroundings seemed to be more crime-sensitive and more positive, with posted text with higher sentiment than anyone else.
The contrary was true for the poorest neighborhoods, where those with the most equal surroundings were less sensitive to crime, and their text had the lowest sentiment of all. Finally, we showed that the content generated by the users can predict their income and inequality levels. We trained multiple machine learning models with features extracted from Nextdoor and predicted the income ($R^2=0.841$) and inequality ($R^2=0.77$) of neighborhoods.
Our work demonstrated that inferring socieconomic factors from the text posted by users is feasible, presenting opportunities both for scientists and policymakers as well as for algorithmic surveillance.

\section*{Appendix}
\begin{table}[!h]
\centering
\small
\begin{adjustbox}{max width=\columnwidth}
\begin{tabular}{|l|l|} 
\hline
\multicolumn{1}{|c|}{\textbf{Official Crime Category}} & \multicolumn{1}{c|}{\textbf{Aggregated Crime Category}}  \\ 
\hline
Anti-social Behaviour                                     & \multirow{3}{*}{Drugs and Order}                    \\ 
\cline{1-1}
Drugs                                                     &                                                     \\ 
\cline{1-1}
Public Order                                              &                                                     \\ 
\hline
Bike Theft                                                & \multirow{7}{*}{Theft and Property Damage}          \\ 
\cline{1-1}
Burglary                                                  &                                                     \\ 
\cline{1-1}
Criminal Damage and
  Arson                               &                                                     \\ 
\cline{1-1}
Robbery                                                   &                                                     \\ 
\cline{1-1}
Shoplifting                                               &                                                     \\ 
\cline{1-1}
Theft from the Person                                     &                                                     \\ 
\cline{1-1}
Vehicle Theft                                             &                                                     \\ 
\hline
Violent Crimes                                            & \multirow{3}{*}{Weapons and Violent Crimes}         \\ 
\cline{1-1}
Murder                                                    &                                                     \\ 
\cline{1-1}
Weapons                                                   &                                                     \\
\hline
\end{tabular}
\end{adjustbox}
\caption{Crime categories from official crime databases.}
\label{tab:crime_categories}
\end{table}

\begin{figure}[H]
	\begin{center}
    \includegraphics[max width=0.85\columnwidth]{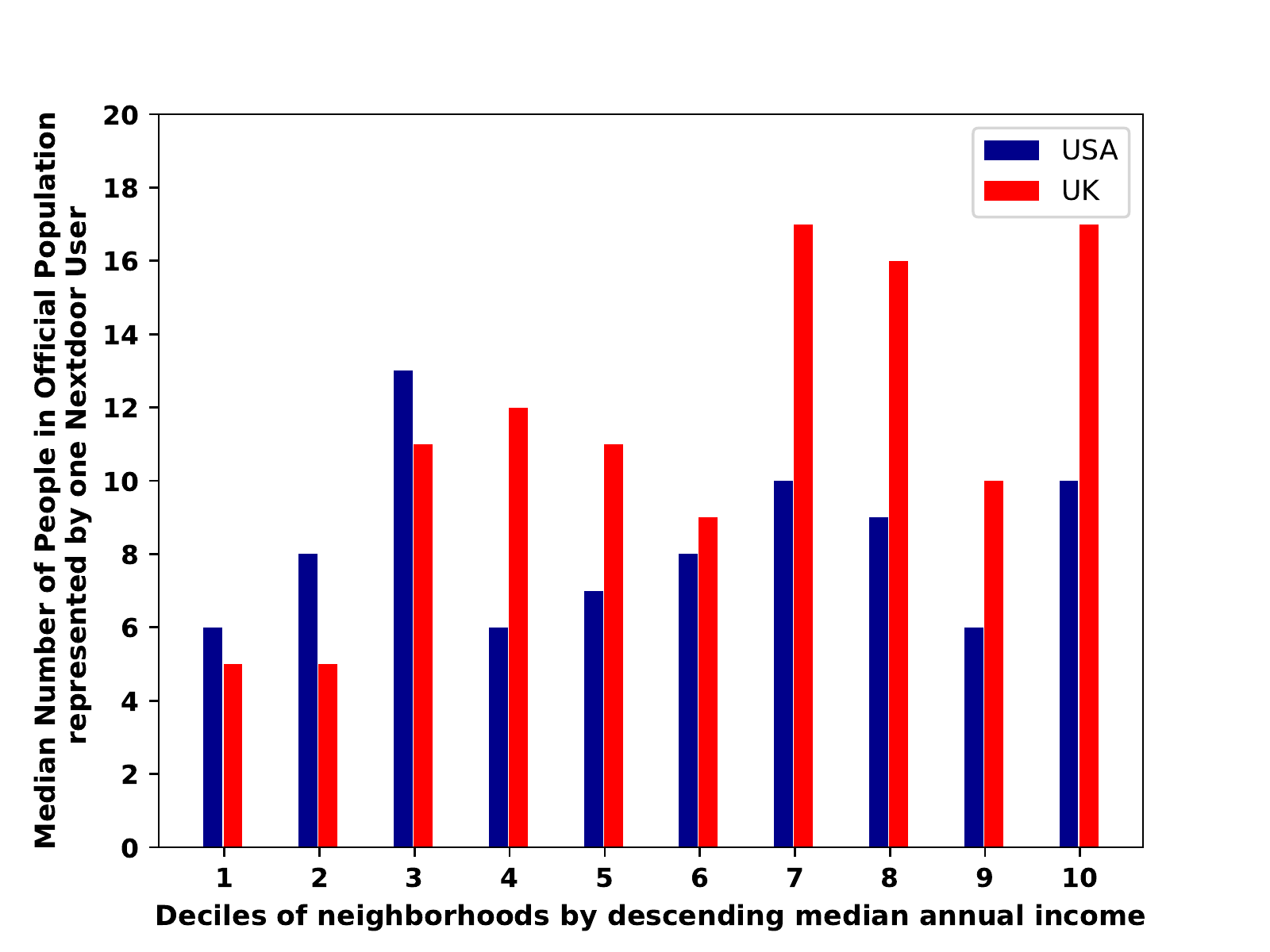}
	\end{center}
	\caption{Population-to-Nextdoor neighbor ratio over  income deciles (from richest to poorest).}
	\label{fig:nextdoor-one-official}
\end{figure}

\begin{figure}[!h]
	\begin{center}
    \includegraphics[max width=0.9\columnwidth]{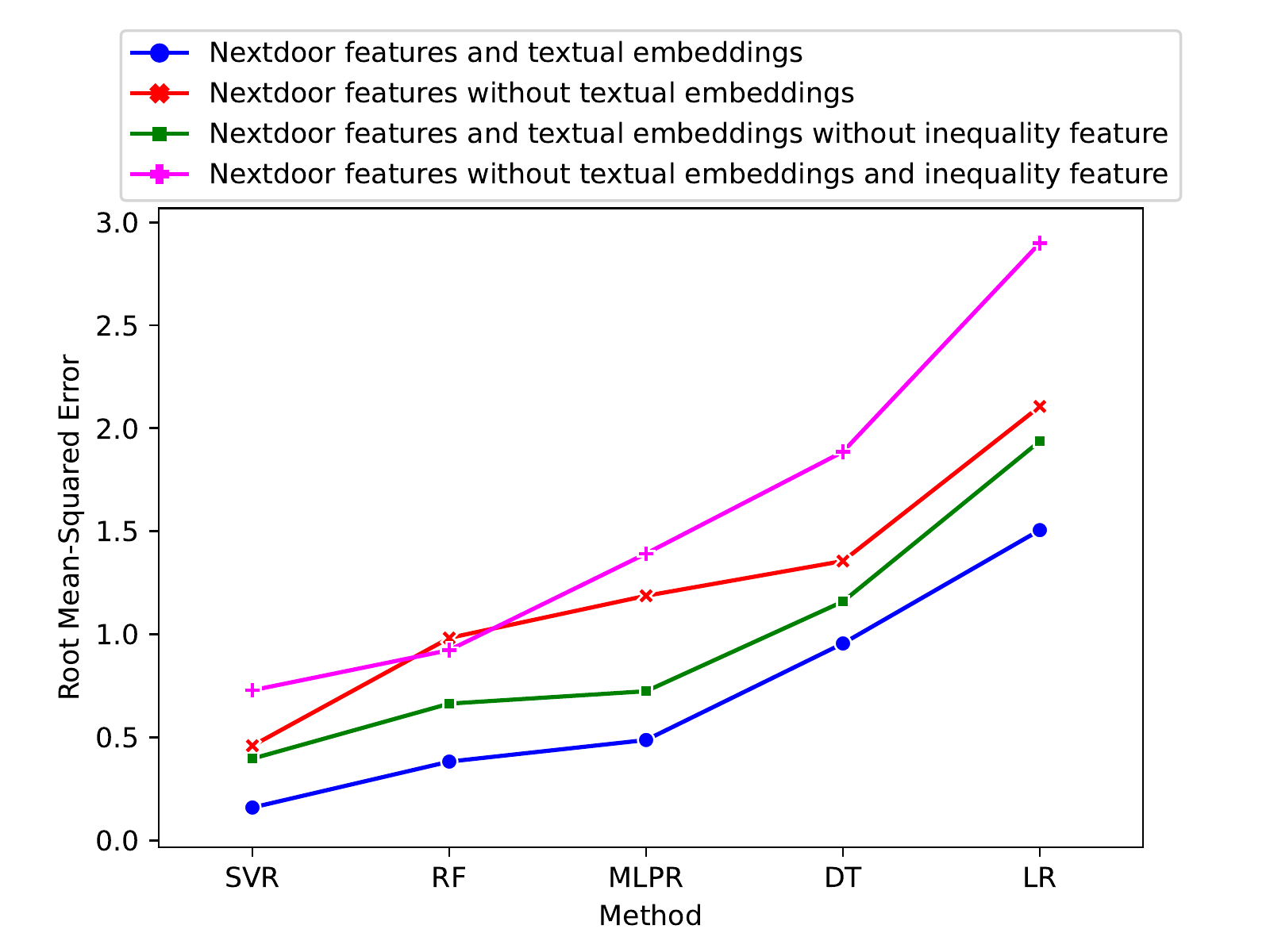}
	\end{center}
	\caption{RMSE values for regression methods on Nextdoor and Official data features to predict income.}
	\label{fig:rmse}
\end{figure}

\begin{figure}[!h]
	\begin{center}
    \includegraphics[width=0.9\columnwidth]{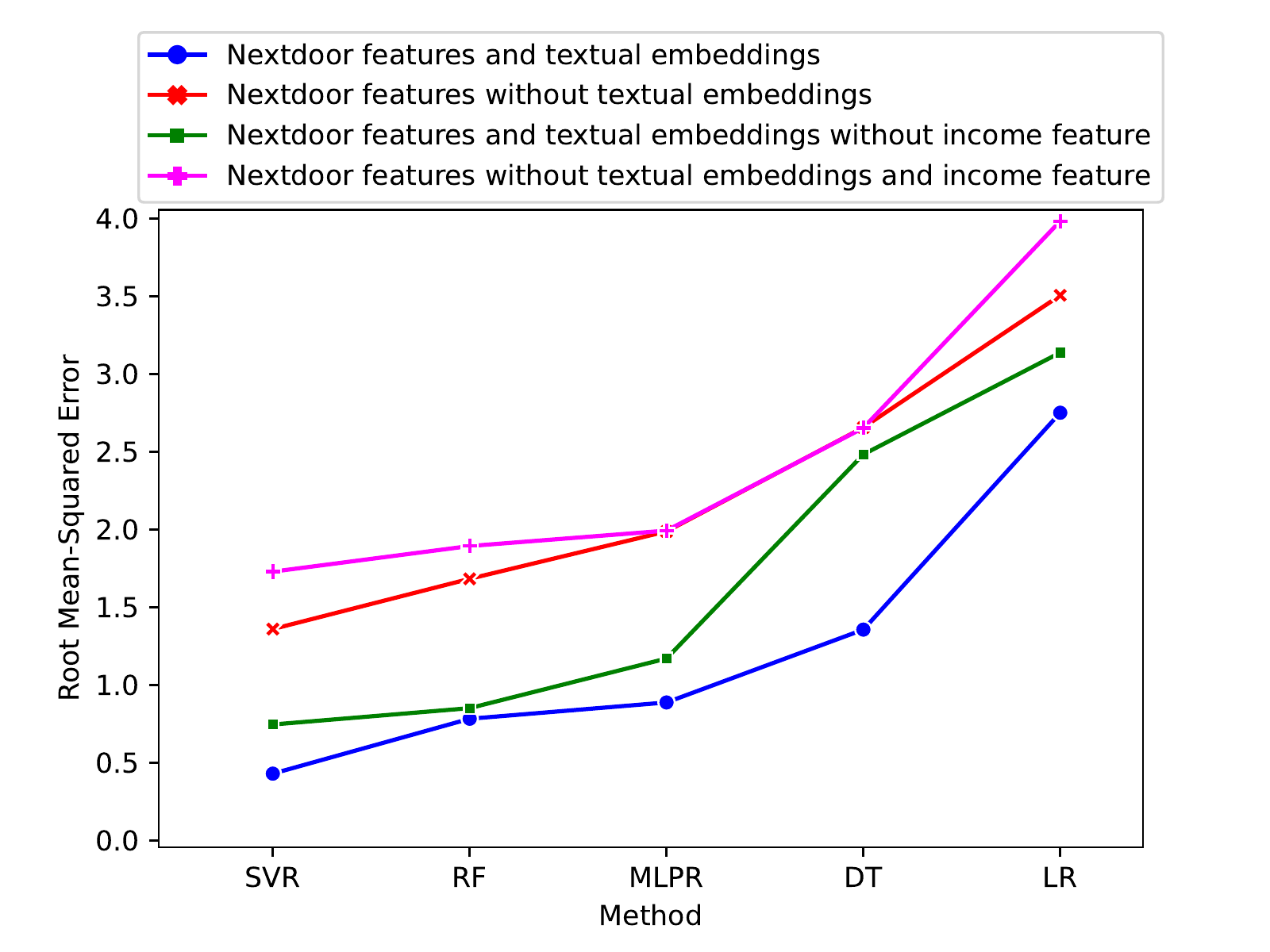}
	\end{center}
	\caption{RMSE values for regression methods on Nextdoor data and Official data features to predict income inequality.}
	\label{fig:rmse_inequality}
\end{figure}
\begin{table}[!h]
\centering
\normalsize
\begin{adjustbox}{max width=\linewidth}
\begin{tabular}{|l|l|} 
\hline
\textbf{Model}    & \textbf{Model Parameters}                                                                                                                                                                                                                                                                                                                                                           \\ 
\hline
SVR               & \begin{tabular}[c]{@{}l@{}l@{}l@{}}C:5, kernel: rbf, degree: 3, gamma: scale \\(income)\\\hline C:10, kernel: rbf, degree: 2, gamma: scale \\(inequality)\\\hline C:2, kernel: rbf, degree: 1, gamma: auto \\(income without inequality)\\\hline C:3, kernel: rbf, degree: 1, gamma: scale \\(inequality without income)\end{tabular}                                                                                                                                                                                                                                 \\ 
\hline
Random Forest     & \begin{tabular}[c]{@{}l@{}l@{}l@{}}max\_depth:~3, max\_features: sqrt, n\_estimators: 20 \\(income)\\\hline max\_depth: 4, max\_features: auto, n\_estimators: 80 \\(inequality)\\\hline max\_depth: 5, max\_features: auto, n\_estimators: 150 \\(income without inequality)\\\hline max\_depth:~5, max\_features: sqrt, n\_estimators: 60 \\(inequality without income)\\\end{tabular}                              
\\ 
\hline
MLPR              & \begin{tabular}[c]{@{}l@{}}hidden\_layer\_sizes': (120,80,40), max\_iter: 5000, \\activation: relu, solver: adam, alpha: 0.01, \\learning\_rate: adaptive (income)\\\hline hidden\_layer\_sizes': (120,80,40), max\_iter: 5000, \\activation: relu, solver: adam, alpha: 0.01, \\learning\_rate: adaptive (inequality)\\\hline hidden\_layer\_sizes': (150,100,50), max\_iter: 5000, \\activation: tanh, solver: sgd, alpha: 0.05, \\learning\_rate: adaptive (income without inequality)\\\hline hidden\_layer\_sizes': (100,50,30), max\_iter: 5000, \\activation: tanh, solver: sgd, alpha: 0.01, \\learning\_rate: adaptive (inequality without income)\end{tabular}  \\ 
\hline
Decision Trees     & \begin{tabular}[c]{@{}l@{}l@{}l@{}}splitter: best, max\_depth: 7, min\_leaf\_sample: 0.4, \\max\_leaf\_node: 50, max\_features: log2 (income)\\\hline splitter: best, max\_depth: 5, min\_leaf\_sample: 0.3, \\max\_leaf\_node: 40, max\_features: log2 (inequality)\\\hline splitter: best, max\_depth: 7, min\_leaf\_sample: 0.1, \\max\_leaf\_node: 7, max\_features: log2 \\(income without inequality)\\\hline splitter: best, max\_depth: 3, min\_leaf\_sample: 0.2, \\max\_leaf\_node: 40, max\_features: auto \\(inequality without income)\end{tabular}                                                                                                                  \\ 
\hline
Linear Regression & \begin{tabular}[c]{@{}l@{}}alpha=0.6, fit\_intercept=True, max\_iter=2000, \\tol=0.0005, selection='random' \\(income)\\\hline alpha=0.5, fit\_intercept=True, max\_iter=2000, \\tol=0.0005, selection='random' \\(inequality)\\\hline alpha=0.1, fit\_intercept=True, max\_iter=5000, \\tol=0.001, selection='random' \\(income without inequality)\\\hline alpha=0.5, fit\_intercept=True, max\_iter=4000, \\tol=0.003, selection='random' \\(inequality without income)\end{tabular}                                                                                                                                                    \\
\hline
\end{tabular}
\end{adjustbox}
\caption{Best parameters for each model used to predict income and income inequality.}
\label{tab:model_parameters}
\end{table}
\section*{Ethics} 
\label{sec:ethics}
This research study has been approved by the Institutional Review Board (IRB) at the researchers’ institution.
The authors have no competing interests or funding that could undermine this research.
We employ users' public post records from Nextdoor to study their conversations. 
Nextdoor data is public, as there is the  expectation that strangers can view the posts~\cite{townsend2016social}.
Upon collection, we anonymize the data before use and store it in a secure silo.
To prevent user identification, we aggregate our data and analyse at a neighborhood level. After aggregation, we discard any user-level information. 
Our work does not share or redistribute Nextdoor content, as per Nextdoor's Terms of Service. Importantly, web crawling is legal for non-commercial research in the UK \cite{scraping-uk} and the USA \cite{scraping-usa}, where the data collection is performed.

From a \textbf{broader perspective}, our analysis demonstrates that predicting the income of users based on their online discourse is feasible. We believe this finding is an important contribution, particularly as this might further enable algorithmic surveillance~\cite{zuboff2015big} by making easier to segment users based on their economic circumstances.

\section*{Acknowledgements}
This work is supported by EPSRC REPHRAIN ``Moderation in Decentralised Social Networks" (EP/V011189/1), 
 Sodestream (EP/S033564/1), AP4L (EP/W032473/1), 
 COMET project (TED2021-132900A-I00), funded by MCIN/AEI /10.13039/501100011033 and the European Union-NextGenerationEU/PRTR, and 
 REACT-COMODIN-CM-23459 funded by the Comunidad de Madrid and the European Regional Development Fund (ERDF). 
 Guillermo Suarez-Tangil was partially funded by the ``Ramon y Cajal'' Fellowship RYC-2020-029401-I funded by MCIN/AEI/10.13039/501100011033 and ESF ``The European Social Fund invests in your future''.
 \bibliography{aaai23}   

\begin{thebibliography}{46}
\providecommand{\natexlab}[1]{#1}

\bibitem[{Abitbol, Karsai, and Fleury(2018)}]{abitbol2018location}
Abitbol, J.~L.; Karsai, M.; and Fleury, E. 2018.
\newblock Location, occupation, and semantics based socioeconomic status
  inference on twitter.
\newblock In \emph{2018 IEEE International Conference on Data Mining Workshops
  (ICDMW)}, 1192--1199. IEEE.

\bibitem[{Aggarwal, Almeida, and Kumaraguru(2013)}]{aggarwal2013detection}
Aggarwal, A.; Almeida, J.; and Kumaraguru, P. 2013.
\newblock Detection of spam tipping behaviour on foursquare.
\newblock In \emph{Proceedings of the 22nd International Conference on World
  Wide Web}, 641--648.

\bibitem[{Aggarwal, Hinneburg, and Keim(2001)}]{Aggarwal2001}
Aggarwal, C.~C.; Hinneburg, A.; and Keim, D.~A. 2001.
\newblock On the Surprising Behavior of Distance Metrics in High Dimensional
  Space.
\newblock In Van~den Bussche, J.; and Vianu, V., eds., \emph{Database Theory
  --- ICDT 2001}, 420--434. Berlin, Heidelberg: Springer Berlin Heidelberg.

\bibitem[{Aletras and Chamberlain(2018)}]{aletras2018predicting}
Aletras, N.; and Chamberlain, B.~P. 2018.
\newblock Predicting twitter user socioeconomic attributes with network and
  language information.
\newblock In \emph{Proceedings of the 29th on Hypertext and Social Media},
  20--24.

\bibitem[{Arora, Liang, and Ma(2017)}]{Arora2017}
Arora, S.; Liang, Y.; and Ma, T. 2017.
\newblock A Simple but Tough-to-Beat Baseline for Sentence Embeddings.
\newblock In \emph{ICLR}.

\bibitem[{Atkinson, Micklewright, and
  Micklewright(1992)}]{atkinson1992economic}
Atkinson, A.~B.; Micklewright, J.; and Micklewright, M. 1992.
\newblock \emph{Economic transformation in Eastern Europe and the distribution
  of income}.
\newblock Cambridge University Press.

\bibitem[{Bernstein(1960)}]{bernstein1960language}
Bernstein, B. 1960.
\newblock Language and social class.
\newblock \emph{The British journal of sociology}, 11(3): 271--276.

\bibitem[{Bin~Zia et~al.(2022)Bin~Zia, Raman, Castro, Hassan~Anaobi,
  De~Cristofaro, Sastry, and Tyson}]{bin2022toxicity}
Bin~Zia, H.; Raman, A.; Castro, I.; Hassan~Anaobi, I.; De~Cristofaro, E.;
  Sastry, N.; and Tyson, G. 2022.
\newblock Toxicity in the decentralized web and the potential for model
  sharing.
\newblock \emph{Proceedings of the ACM on Measurement and Analysis of Computing
  Systems}, 6(2): 1--25.

\bibitem[{Breiman et~al.(2017)Breiman, Friedman, Olshen, and
  Stone}]{breiman2017classification}
Breiman, L.; Friedman, J.~H.; Olshen, R.~A.; and Stone, C.~J. 2017.
\newblock \emph{Classification and regression trees}.
\newblock Routledge.

\bibitem[{Census(2022)}]{census-usa}
Census, U. 2022.
\newblock \textit{US Census Bureau Release}.
\newblock https://www.census.gov/data.html.
\newblock Accessed: 2022-07-10.

\bibitem[{Chen et~al.(2020)Chen, Dewi, Huang, and Caraka}]{chen2020selecting}
Chen, R.-C.; Dewi, C.; Huang, S.-W.; and Caraka, R.~E. 2020.
\newblock Selecting critical features for data classification based on machine
  learning methods.
\newblock \emph{Journal of Big Data}, 7(1): 1--26.

\bibitem[{Chorley et~al.(2016)Chorley, Rossi, Tyson, and
  Williams}]{chorley2016pub}
Chorley, M.~J.; Rossi, L.; Tyson, G.; and Williams, M.~J. 2016.
\newblock Pub crawling at scale: tapping untappd to explore social drinking.
\newblock In \emph{Tenth International AAAI Conference on Web and Social
  Media}.

\bibitem[{Cui et~al.(2022)Cui, Zhang, Jaidka, Pang, Sherman, Jakhetiya, Ungar,
  and Guntuku}]{cui2022social}
Cui, J.; Zhang, T.; Jaidka, K.; Pang, D.; Sherman, G.; Jakhetiya, V.; Ungar,
  L.~H.; and Guntuku, S.~C. 2022.
\newblock Social Media Reveals Urban-Rural Differences in Stress across China.
\newblock In \emph{Proceedings of the International AAAI Conference on Web and
  Social Media}, volume~16, 114--124.

\bibitem[{Dekker(2007)}]{dekker2007social}
Dekker, K. 2007.
\newblock Social capital, neighbourhood attachment and participation in
  distressed urban areas. A case study in The Hague and Utrecht, the
  Netherlands.
\newblock \emph{Housing Studies}, 22(3): 355--379.

\bibitem[{Easterlin(1974)}]{easterlin1974does}
Easterlin, R.~A. 1974.
\newblock Does economic growth improve the human lot? Some empirical evidence.
\newblock In \emph{Nations and households in economic growth}, 89--125.
  Elsevier.

\bibitem[{FBI(2022)}]{fbi-usa}
FBI. 2022.
\newblock \textit{Federal Bureau of Investigation Crime Data Explorer}.
\newblock https://cde.ucr.cjis.gov/LATEST/webapp/pages/home.
\newblock Accessed: 2022-07-10.

\bibitem[{GeoPy(2022)}]{geopy-python}
GeoPy. 2022.
\newblock \textit{GeoPy}.
\newblock https://geopy.readthedocs.io/en/stable/.
\newblock Accessed: 2022-07-10.

\bibitem[{Giorgi et~al.(2021)Giorgi, Guntuku, Eichstaedt, Pajot, Schwartz, and
  Ungar}]{giorgi2021well}
Giorgi, S.; Guntuku, S.~C.; Eichstaedt, J.~C.; Pajot, C.; Schwartz, H.~A.; and
  Ungar, L.~H. 2021.
\newblock Well-Being Depends on Social Comparison: Hierarchical Models of
  Twitter Language Suggest That Richer Neighbors Make You Less Happy.
\newblock In \emph{Proceedings of the International AAAI Conference on Web and
  Social Media}, volume~15, 1069--1074.

\bibitem[{Hays and Kogl(2007)}]{hays2007neighborhood}
Hays, R.~A.; and Kogl, A.~M. 2007.
\newblock Neighborhood attachment, social capital building, and political
  participation: A case study of low-and moderate-income residents of Waterloo,
  Iowa.
\newblock \emph{Journal of Urban Affairs}, 29(2): 181--205.

\bibitem[{Huggingface(2022)}]{huggingface}
Huggingface. 2022.
\newblock \textit{all-mpnet-base-v2}.
\newblock https://huggingface.co/sentence-transformers/all-mpnet-base-v2.
\newblock Accessed: 2022-07-10.

\bibitem[{Hutto and Gilbert(2014)}]{hutto2014vader}
Hutto, C.; and Gilbert, E. 2014.
\newblock Vader: A parsimonious rule-based model for sentiment analysis of
  social media text.
\newblock In \emph{Proceedings of the international AAAI conference on web and
  social media}, volume~8, 216--225.

\bibitem[{Inman(1849)}]{inman1849navigation}
Inman, J. 1849.
\newblock \emph{Navigation and nautical astronomy, for the Use of British
  Seamen}.
\newblock F. \& J. Rivington.

\bibitem[{IPO(2021)}]{scraping-uk}
IPO, U. 2021.
\newblock \textit{Exceptions to copyright}.
\newblock
  https://www.gov.uk/guidance/exceptions-to-copyright\#text-and-data-mining-for-non-commercial-research.
\newblock Accessed: 2022-07-10.

\bibitem[{Iqbal et~al.(2022)Iqbal, Arshad, Tyson, and
  Castro}]{iqbal2022exploring}
Iqbal, W.; Arshad, M.~H.; Tyson, G.; and Castro, I. 2022.
\newblock Exploring Crowdsourced Content Moderation Through Lens of Reddit
  during COVID-19.
\newblock In \emph{Proceedings of the 17th Asian Internet Engineering
  Conference}, 26--35.

\bibitem[{Jimenez~Villalonga(2021)}]{jimenez2021uncovering}
Jimenez~Villalonga, F. 2021.
\newblock Uncovering Correlations Between Two UMAP Hyperparameters and the
  Input Dataset.
\newblock 1: 14.

\bibitem[{Lampos et~al.(2016)Lampos, Aletras, Geyti, Zou, and
  Cox}]{lampos2016inferring}
Lampos, V.; Aletras, N.; Geyti, J.~K.; Zou, B.; and Cox, I.~J. 2016.
\newblock Inferring the socioeconomic status of social media users based on
  behaviour and language.
\newblock In \emph{European conference on information retrieval}, 689--695.
  Springer.

\bibitem[{Lee~Rodgers and Nicewander(1988)}]{lee1988thirteen}
Lee~Rodgers, J.; and Nicewander, W.~A. 1988.
\newblock Thirteen ways to look at the correlation coefficient.
\newblock \emph{The American Statistician}, 42(1): 59--66.

\bibitem[{McHugh(2012)}]{mchugh2012interrater}
McHugh, M.~L. 2012.
\newblock Interrater reliability: the kappa statistic.
\newblock \emph{Biochemia medica}, 22(3): 276--282.

\bibitem[{McInnes et~al.(2018)McInnes, Healy, Saul, and
  Gro{\ss}berger}]{McInnes2018}
McInnes, L.; Healy, J.; Saul, N.; and Gro{\ss}berger, L. 2018.
\newblock UMAP: Uniform Manifold Approximation and Projection.
\newblock \emph{Journal of Open Source Software}, 3(29).

\bibitem[{Moro et~al.(2021)Moro, Calacci, Dong, and
  Pentland}]{moro2021mobility}
Moro, E.; Calacci, D.; Dong, X.; and Pentland, A. 2021.
\newblock Mobility patterns are associated with experienced income segregation
  in large US cities.
\newblock \emph{Nature communications}, 12(1): 1--10.

\bibitem[{Nextdoor(2021)}]{nextdoor-stats}
Nextdoor. 2021.
\newblock \textit{Nextdoor can bring communities together}.
\newblock
  https://inews.co.uk/inews-lifestyle/nextdoor-app-neighbourhood-online-groups-780656.
\newblock Accessed: 2022-07-10.

\bibitem[{Nextdoor(2022)}]{nextdoor-neighborhoods}
Nextdoor. 2022.
\newblock \textit{Nextdoor Neighborhoods}.
\newblock https://nextdoor.com/find-neighborhood/.
\newblock Accessed: 2022-07-10.

\bibitem[{ONS(2018)}]{lsoa-uk}
ONS, U. 2018.
\newblock \textit{Lower Layer Super Output Area (2001) to Lower Layer Super
  Output Area (2011) to Local Authority District (2011) Lookup in England and
  Wales}.
\newblock
  https://www.data.gov.uk/dataset/afc2ed54-f1c5-44f3-b8bb-6454eb0153d0/lower-layer-super-output-area-2001-to-lower-layer-super-output-area-2011-to-local-authority-district-2011-lookup-in-england-and-wales.
\newblock Accessed: 2022-07-10.

\bibitem[{Police(2022)}]{met-uk}
Police, U.~M. 2022.
\newblock \textit{Stats and data}.
\newblock https://www.met.police.uk/sd/stats-and-data/.
\newblock Accessed: 2022-07-10.

\bibitem[{Preo{\c{t}}iuc-Pietro, Lampos, and
  Aletras(2015)}]{preoctiuc2015analysis}
Preo{\c{t}}iuc-Pietro, D.; Lampos, V.; and Aletras, N. 2015.
\newblock An analysis of the user occupational class through Twitter content.
\newblock In \emph{Proceedings of the 53rd Annual Meeting of the Association
  for Computational Linguistics and the 7th International Joint Conference on
  Natural Language Processing (Volume 1: Long Papers)}, 1754--1764.

\bibitem[{Preo{\c{t}}iuc-Pietro et~al.(2015)Preo{\c{t}}iuc-Pietro, Volkova,
  Lampos, Bachrach, and Aletras}]{preoctiuc2015studying}
Preo{\c{t}}iuc-Pietro, D.; Volkova, S.; Lampos, V.; Bachrach, Y.; and Aletras,
  N. 2015.
\newblock Studying user income through language, behaviour and affect in social
  media.
\newblock \emph{PloS one}, 10(9): e0138717.

\bibitem[{Reimers and Gurevych(2019)}]{reimers-2019-sentence-bert}
Reimers, N.; and Gurevych, I. 2019.
\newblock Sentence-BERT: Sentence Embeddings using Siamese BERT-Networks.
\newblock In \emph{Proceedings of the 2019 Conference on Empirical Methods in
  Natural Language Processing}. Association for Computational Linguistics.

\bibitem[{TechCrunch(2022)}]{scraping-usa}
TechCrunch. 2022.
\newblock \textit{Web scraping is legal, US appeals court reaffirms}.
\newblock https://techcrunch.com/2022/04/18/web-scraping-legal-court/.
\newblock Accessed: 2022-07-10.

\bibitem[{T{\'o}th et~al.(2021)T{\'o}th, Wachs, Di~Clemente, Jakobi,
  S{\'a}gv{\'a}ri, Kert{\'e}sz, and Lengyel}]{toth2021inequality}
T{\'o}th, G.; Wachs, J.; Di~Clemente, R.; Jakobi, {\'A}.; S{\'a}gv{\'a}ri, B.;
  Kert{\'e}sz, J.; and Lengyel, B. 2021.
\newblock Inequality is rising where social network segregation interacts with
  urban topology.
\newblock \emph{Nature communications}, 12(1): 1--9.

\bibitem[{Townsend and Wallace(2016)}]{townsend2016social}
Townsend, L.; and Wallace, C. 2016.
\newblock Social media research: A guide to ethics.
\newblock \emph{University of Aberdeen}, 1: 16.

\bibitem[{UK(2022)}]{population-uk}
UK, O. 2022.
\newblock \textit{Estimates of the population for the UK, England, Wales,
  Scotland and Northern Ireland}.
\newblock
  https://www.ons.gov.uk/peoplepopulationandcommunit/\\populationandmigration/populationestimates/dataset/\\populationestimatesforukenglandandwalesscotlandand//\\northernireland.
\newblock Accessed: 2022-07-10.

\bibitem[{Wu(2012)}]{wu2012neighborhood}
Wu, F. 2012.
\newblock Neighborhood attachment, social participation, and willingness to
  stay in China’s low-income communities.
\newblock \emph{Urban Affairs Review}, 48(4): 547--570.

\bibitem[{Xu et~al.(2019)Xu, Belyi, Santi, and Ratti}]{xu2019quantifying}
Xu, Y.; Belyi, A.; Santi, P.; and Ratti, C. 2019.
\newblock Quantifying segregation in an integrated urban physical-social space.
\newblock \emph{Journal of the Royal Society Interface}, 16(160): 20190536.

\bibitem[{Zhang et~al.(2022)Zhang, Fang, Chen, and
  Namazi-Rad}]{zhang2022neural}
Zhang, Z.; Fang, M.; Chen, L.; and Namazi-Rad, M.-R. 2022.
\newblock Is Neural Topic Modelling Better than Clustering? An Empirical Study
  on Clustering with Contextual Embeddings for Topics.
\newblock \emph{arXiv preprint arXiv:2204.09874}.

\bibitem[{Zia et~al.(2023)Zia, He, Raman, Castro, Sastry, and
  Tyson}]{zia2023flocking}
Zia, H.~B.; He, J.; Raman, A.; Castro, I.; Sastry, N.; and Tyson, G. 2023.
\newblock Flocking to mastodon: Tracking the great twitter migration.
\newblock \emph{arXiv preprint arXiv:2302.14294}.

\bibitem[{Zuboff(2015)}]{zuboff2015big}
Zuboff, S. 2015.
\newblock Big other: surveillance capitalism and the prospects of an
  information civilization.
\newblock \emph{Journal of information technology}, 30(1): 75--89.

\end{thebibliography}
\end{document}